\documentclass[a4paper,twocolumn,11pt,accepted=2023-07-15]{quantumarticle}
\pdfoutput=1
\usepackage[utf8]{inputenc}
\usepackage[english]{babel}
\usepackage[T1]{fontenc}
\usepackage[numbers,sort&compress]{natbib}
\usepackage{amsmath}
\usepackage{physics}
\usepackage{hyperref}
\usepackage{tikz}
\usepackage{lipsum}
\usepackage{multirow}
\usepackage{algorithm}
\usepackage[noend]{algpseudocode}
\newtheorem{theorem}{Theorem}
\renewcommand{\Re}{\mathrm{Re}}
\renewcommand{\Im}{\mathrm{Im}}

\begin{document}

\title{Quantum-assisted Monte Carlo algorithms for fermions}

\author{Xiaosi Xu}
\affiliation{Graduate School of China Academy of Engineering Physics, Beijing 100193, China}
\orcid{0000-0002-4894-8322}

\author{Ying Li}
\email{yli@gscaep.ac.cn}
\affiliation{Graduate School of China Academy of Engineering Physics, Beijing 100193, China}
\orcid{0000-0002-1705-2494}

\maketitle

\begin{abstract}
Quantum computing is a promising way to systematically solve the longstanding computational problem, the ground state of a many-body fermion system. Many efforts have been made to realise certain forms of quantum advantage in this problem, for instance, the development of variational quantum algorithms. A recent work by Huggins et al.~\cite{Huggins2022} reports a novel candidate, i.e.~a quantum-classical hybrid Monte Carlo algorithm with a reduced bias in comparison to its fully-classical counterpart. In this paper, we propose a family of scalable quantum-assisted Monte Carlo algorithms where the quantum computer is used at its minimal cost and still can reduce the bias. By incorporating a Bayesian inference approach, we can achieve this quantum-facilitated bias reduction with a much smaller quantum-computing cost than taking empirical mean in amplitude estimation. Besides, we show that the hybrid Monte Carlo framework is a general way to suppress errors in the ground state obtained from classical algorithms. Our work provides a Monte Carlo toolkit for achieving quantum-enhanced calculation of fermion systems on near-term quantum devices. 
\end{abstract}

\section{Introduction}

The many-body Schr\"{o}dinger equation is a longstanding computational challenge in quantum chemistry~\cite{babbush2016exponentially,mcardle2020quantum}, condensed matter physics~\cite{resta2000manifestations,guo2020condensed}, nuclear physics~\cite{jeukenne1976many,carlson2015quantum}, quantum chromodynamics~\cite{miransky2015quantum,brodsky1998quantum}, etc. Over the years, many numerical methods are developed to address this problem, for instance, mean-field theory~\cite{kotliar2006electronic, negele1982mean}, Monte Carlo method~\cite{guardiola1998monte} and tensor-network formalism~\cite{shi2006classical,ran2017few}. These methods have played essential roles in basic science and have also been applied to various fields outside physics, such as economic modelling~\cite{ creal2012survey, batan2016techno} and machine learning~\cite{ sun2020generative, tanaka1998mean}. However, because of the infamous difficulty that the Hilbert space dimension grows exponentially with the particle number, all known methods are limited to small-size systems, approximate solutions or specific models. 

Among conventional numerical methods, quantum Monte Carlo (QMC) is a group of classical algorithms that can bypass the difficulty of exponentially growing space dimensions~\cite{austin2012quantum}. Specifically, QMC applies probabilistic sampling in a subset (e.g. Slater determinants) of the entire Hilbert space, resulting in a polynomial scaling of the memory size. However, for a generic Hamiltonian, QMC suffers from the sign problem where the sign of the sampling integrands oscillates alternatively from positive to negative~\cite{ortiz2001quantum}. The sign problem can lead to an impermissible statistical error. Commonly used methods to constrain the sign problem, like the phaseless approximation~\cite{motta2018ab} and fixed-node approximation~\cite{Blunt2021}, usually bias the result. The bias is largely determined by the so-called trial state. 

In recent years, quantum computing has been experiencing fast development and has shown the potential to solve classically intractable problems. Quantum computing offers a scalable way to solve the many-body Schr\"{o}dinger equation given a good guiding wavefunction~\cite{gharibian2022dequantizing,cade2022complexity,gharibian2022improved}. An established quantum-computing (QC) approach combines Hamiltonian simulation and quantum phase estimation~\cite{ whitfield2011simulation, cruz2020optimizing}. However, this approach requires fault-tolerant quantum technologies, which are still years away. At present, much attention has been put into quantum algorithms suitable for noisy intermediate-scale quantum (NISQ) devices~\cite{preskill2018quantum}. The variational quantum eigensolver (VQE) and its variants~\cite{bharti2022noisy}, as the currently most popular near-term algorithms, have proved successful in applications of a wide area. Despite this, VQE has challenges such as barren plateaus~\cite{wang2021noise,cerezo2021cost}, which motivates the recent development of optimiser and variational quantum circuit design~\cite{grant2019initialization,sack2022avoiding}. Meanwhile, feasible and non-variational approaches with potential quantum advantages are urged. 

A promising hybrid approach is quantum computing incorporating QMC. Recently, Huggins et al. proposed a new way to calculate the ground state of fermion systems, a quantum-classical hybrid auxiliary-field Monte Carlo (AFMC) algorithm, with an experimental 
demonstration that it can reduce the bias~\cite{Huggins2022}. The algorithm utilises a trial state prepared on a quantum computer replacing the conventional trial state. An alternative way to incorporate QMC is sampling states not from a subset in favour of classical computing but those generated by quantum circuits~\cite{yang2021accelerated}. In this way, one can simulate the time evolution with a trade-off between the sign problem and the quantum circuit complexity. Either way, one of the central issues is whether a quantum advantage, i.e.~a smaller bias or sample size compared with classical algorithms, is attainable on NISQ devices. 

This paper presents a framework of assisting QMC algorithms with states prepared on a quantum computer. We generalise the original work of Huggins et al.~\cite{Huggins2022} by utilizing quantum computing at flexible levels, and the QC state is used in various ways other than guiding the QMC sampling. Although shadow tomography has been proposed as the interface that passes the state from a quantum computer to a classical computer, we focus on the Hadamard test as an alternative way of measuring amplitudes of the QC state. We notice that the idea of flexible-level quantum-assisted computing works both ways. One of the purposes of this work is to address the measurement cost issue, which has raised debates about the potential quantum advantage~\cite{mazzola2022exponential,lee2022response}. By carrying out QMC calculation in a bi-trial-state manner, i.e.~taking use of a classical-computing (CC) trial state in addition to a QC trial state, we can reduce the measurement cost required for demonstrating quantum advantage. Here we list the main results as follows: 
\begin{enumerate}
\item We give a framework of quantum-assisted variational Monte Carlo (VMC), Green's function Monte Carlo (GFMC) and AFMC algorithms; 
\item Depending on the extent a quantum computer is involved in computation, we name two strategies, the {\it consistent} quantum-assisted (CQA, or consistent quantum-classical) algorithms and the quantum-assisted energy evaluation (QAEE) algorithms, and the latter use minimal QC resources; 
\item We propose a general Bayesian inference method to reduce quantum measurements and achieve a stable quantum advantage; 
\item We show that quantum-assisted QMC is error-resilient due to an inherent symmetry projection; 
\item Lastly, we propose the quantum-classical Monte Carlo subspace diagonalisation (QCMCSD) algorithm and show that hybrid QMC is a general way to reduce errors in state-of-the-art classical algorithms. 
\end{enumerate}

In QMC algorithms, evaluating amplitudes of the trial state is the elementary operation. Each amplitude evaluation on a quantum computer requires a number of quantum measurements (which are called circuit shots in this paper). When the circuit shot number is finite, the amplitude evaluation has a statistical error. In what follows, we use QC variance to denote this error source. We can reduce the impact of this error in various ways. In CQA algorithms, the output energy is variational in the sense that it is always higher than the true ground-state energy, even if the amplitude evaluation has a significant statistical error. Because of the variational property, we can enhance the quantum advantage by post-selection, i.e.~we run the algorithm a few times and select the lowest energy. Using the Bayesian inference method, we find that a bias smaller than the classical algorithm occurs with a high probability even with a small circuit shot number. In QAEE algorithms, we only utilise the quantum computer to evaluate the local energy at the last stage, such that the cost of circuit shots is minimised.

This paper is structured as follows. In section~\ref{sec:algorithms_basic} we give an introduction to the Monte Carlo methods related to this work. We present these methods in a way appropriate for the generalisation to quantum computing. The main algorithms and numerical results are presented in section~\ref{sec:algorithm}. In section~\ref{sec:amplitude} we introduce three scalable methods of evaluating amplitudes on a quantum computer. Section~\ref{sec:Bayesian} introduces a method based on the Bayesian inference that can reduce the variance due to the finite QC resources. Section~\ref{sec:symmetry} discusses the symmetries in fermion systems and how we can make use of symmetries to reduce errors. In section~\ref{sec:subspace_diag}, we show an application of the hybrid QMC approach, i.e.~the QCMCSD algorithm, to obtain a solution closer to the true ground state given optimal solutions from classical and quantum algorithms. Finally, this paper is concluded in section~\ref{sec:conclusion}.

\section{Quantum Monte Carlo algorithms}
\label{sec:algorithms_basic}

In VMC~\cite{Mahajan2019, Roggero2013, Sandvik2007}, GFMC~\cite{Blunt2021, Haaf1995} and AFMC~\cite{motta2018ab, zhang2003quantum} algorithms formalised with the mixed estimator, the ground-state energy $E_g$ is computed according to 
\begin{align}
E = \frac{\bra{\Psi_T}H\ket{\Psi}}{\braket{\Psi_T}{\Psi}}.
\label{eq:E}
\end{align}

Here, $\ket{\Psi}$ is an approximation to the ground state $\ket{\Psi_g}$, and $\ket{\Psi_T}$ is the trial state. If $\ket{\Psi} = \ket{\Psi_g}$, $E = E_g$ regardless of the trial state. The trial state is important in these algorithms as we will show later.

QMC algorithms use Monte Carlo integration/summation to evaluate the properties of a physical system. Specifically, to compute the ground-state energy according to Eq.~(\ref{eq:E}), we generate a linear expression of $\ket{\Psi}$ in the form 

\begin{align}
\ket{\Psi} = \sum_{l = 1}^{N} w_{l} e^{i\theta_{l}} \frac{\ket{\phi_{l}}}{\braket{\Psi_T}{\phi_{l}}},
\label{eq:Psi}
\end{align}

where $l$ is the label of the random {\it walker}, $w_{l}$ and $\theta_{l}$ are the weight and phase of the corresponding state $\phi_{l}$, respectively, and $\phi_{l}$ is generated in a stochastic process. On a classical computer, the memory size for storing a generic quantum state usually increases exponentially with the number of particles. In QMC algorithms, $\phi$ can be stored without the exponential memory cost, for example in the form of Slater determinants for fermion systems. 
In the following, we call $\phi_{l}$ the walker state. Given the linear expression, the energy becomes 
\begin{align}
E = \frac{\sum_{l = 1}^{N} w_{l} e^{i\theta_{l}} E^{\mathrm{loc}}(\phi_{l})}{\sum_{l = 1}^{N} w_{l} e^{i\theta_{l}}},
\label{eq:EQMC}
\end{align}
where 
\begin{align}
E^{\mathrm{loc}}(\phi) \equiv \frac{\bra{\Psi_T}H\ket{\phi}}{\braket{\Psi_T}{\phi}}
\label{eq:Eloc}
\end{align}
is called the local energy. Without using a quantum computer, we need to take a trial state that quantities $\braket{\Psi_T}{\phi}$ and $\bra{\Psi_T}H\ket{\phi}$ can be efficiently computed on the classical computer. 

In VMC, the walker states $\ket{\phi}$ are taken from an orthonormal basis of the Hilbert space $\mathcal{R} = \{\ket{\mathbf{R}}\}$. For fermion systems, usually we take Slater determinants (SDs) as the basis. A usual way of constructing the trial state is to apply a Jastrow factor to a mean-field wavefunction~\cite{Mahajan2019}, and VMC works for all trial states that the amplitudes $\braket{\Psi_T}{\mathbf{R}}$ can be computed. Coupled-cluster states~\cite{Roggero2013} and tensor-network states~\cite{Sandvik2007} have been investigated for this purpose in the literature. Given the basis, we can decompose the trial state as 
\begin{align}
\ket{\Psi_T} = \sum_{\ket{\mathbf{R}}\in \mathcal{R}} \abs{\braket{\Psi_T}{\mathbf{R}}}^2 \frac{\ket{\mathbf{R}}}{\braket{\Psi_T}{\mathbf{R}}}.
\end{align}
The summation in the decomposition is evaluated using the Monte Carlo method: $N$ random basis states $\{\ket{\mathbf{R}_l}| l=1,2,\ldots,N\}$ are generated according to the probability distribution $\abs{\braket{\Psi_T}{\mathbf{R}}}^2$. Usually the Metropolis-Hastings algorithm or some variant is used in this procedure~\cite{Sabzevari2018}. 

Taking $\phi_l = \mathbf{R}_l$, $w_l = N^{-1}$ and $\theta_l = 0$ in Eq.~(\ref{eq:Psi}), we have 
\begin{align}
\ket{\Psi} = \frac{1}{N}\sum_{l = 1}^{N} \frac{\ket{\mathbf{R}_{l}}}{\braket{\Psi_T}{\mathbf{R}_{l}}}.
\label{eq:PsiVMC}
\end{align}
This state $\ket{\Psi}$ converges to $\ket{\Psi_T}$ in the limit $N\rightarrow\infty$. 
Substituting $\ket{\Psi}$ in Eq.~(\ref{eq:PsiVMC}) for $\ket{\Psi_T}$ (not for $\bra{\Psi_T}$) in Eq.~(\ref{eq:E}), we have the final expression of the energy 
\begin{align}
E = \frac{1}{N} \sum_{l = 1}^{N} E^{\mathrm{loc}}(\mathbf{R}_{l}),
\label{eq:Eave}
\end{align}
which coincides with Eq.~(\ref{eq:EQMC}). Essential steps of the energy evaluation in VMC can be found in Algorithm~\ref{alg:VMC}.

Given the energy evaluation, VMC minimizes the expected energy of a parameterised trial state $\ket{\Psi_T(\boldsymbol{\lambda})}$ variationally, where $\boldsymbol{\lambda}$ denotes parameters. The expected energy is 
\begin{align}
E(\boldsymbol{\lambda}) \equiv \frac{\bra{\Psi_T(\boldsymbol{\lambda})}H\ket{\Psi_T(\boldsymbol{\lambda})}}{\braket{\Psi_T(\boldsymbol{\lambda})}{\Psi_T(\boldsymbol{\lambda})}}.
\label{eq:EVMC}
\end{align}

GFMC is often referred as a projector QMC algorithm. For any initial state $\ket{\Psi_I}$ with a nonzero overlap with the ground state, the final state $e^{-n\Delta\beta H}\ket{\Psi_I}\approx (\openone - \Delta\beta H)^n\ket{\Psi_I}$ converges to the ground state in the limit $n\rightarrow\infty$. Here, $e^{-n\Delta\beta H}$ is the imaginary-time propagator, and $\Delta\beta$ is a real parameter that has to be properly chosen. Similar to VMC, we generate random basis states to represent the final state: First, we generate an initial random basis state; then, it evolves in a stochastic process driven by the operator $\openone - \Delta\beta H$ for $n$ steps. In general, phases $\theta_l$ are not zero in GFMC, and they can cause a large statistical error, which is the notorious sign problem. We can deal with this problem by introducing approximations in fixed-node GFMC (fn-GFMC). In the fixed-node approximation, the Hamiltonian is replaced with an oscillatory-sign-free Hamiltonian $H^{fn}$ (called the fixed-node Hamiltonian) that depends on the trial state. In the approach introduced by van Bemmel {\it et al.}~\cite{Blunt2021, Haaf1995}, $E$ in the limit $N,n\rightarrow\infty$ converges to $E_g^{fn}$, the ground-state energy of $H^{fn}$, and $E_g^{fn}$ is variational, i.e.~it always holds that $E_g^{fn}\geq E_g$. We can optimise the trial state to minimise the difference $E_g^{fn}-E_g$ according to the variational principle. When $\ket{\Psi_T} = \ket{\Psi_g}$, the fixed-node ground-state energy is exact, i.e.~$E_g^{fn} = E_g$. See Appendix~\ref{app:GFMC} for a detailed formalism of GFMC. 

AFMC is another projector QMC algorithm. Similar to GFMC, the ground-state energy is computed by simulating the imaginary-time evolution $e^{-n\Delta\beta H}\ket{\Psi_I}$. In AFMC, $\phi$ states are general SDs instead of basis states, and the trial and initial states are also SDs, e.g.~worked out with the Hartree-Fock (HF) or density functional theory methods. Using the Trotter-Suzuki formula and Hubbard-Stratonovich transformation, we can approximate the imaginary-time propagator with an integral over a product of one-particle propagators. This procedure is crucial because the evolution of general SDs under one-particle propagators is tractable on a classical computer. These one-particle propagators depend on an auxiliary field. To generate $\phi_l$ and the corresponding weights and phases, the auxiliary field is randomly sampled for each time step, and the initial SD evolves driven by corresponding one-particle propagators. For a generic fermion Hamiltonian, AFMC also has the sign problem (more specifically, the phase problem). Using the phaseless approximation, which is a different approach from the fixed-node approximation, we can implement AFMC without an oscillatory sign/phase, and this algorithm is called phaseless AFMC (ph-AFMC)~\cite{zhang2003quantum}. With the phaseless approximation, the energy does not converge to the exact ground-state energy in the limit $N,n\rightarrow\infty$. The bias to the ground-state energy depends on the trial state and vanishes when the trial state is exactly the ground state. See Appendix~\ref{app:AFMC} for a detailed formalism of AFMC. 

In QMC algorithms, there are two sources of errors, bias and variance. In VMC, $E$ in Eq.~(\ref{eq:Eave}) converges to $E(\boldsymbol{\lambda})$ in the limit $N\rightarrow\infty$. The bias is the difference between $E(\boldsymbol{\lambda})$ and the true ground-state energy $E_g$. When $N$ is finite, the Monte Carlo method has a finite accuracy due to the statistical error, i.e.~an error to the true mean $E(\boldsymbol{\lambda})$. Both of them contribute to the final error in $E$. In GFMC and AFMC, we can systematically reduce the bias to zero; however, the variance may increase exponentially with time steps because of the sign problem. We can prevent the sign problem by modifying the exact GFMC and AFMC formalisms and introducing approximations. In this paper, we focus on oscillatory-sign-free GFMC and AFMC algorithms. Specifically, we consider fn-GFMC and ph-AFMC. These algorithms have a residual bias because of approximations. 

To summarise, some features of these algorithms can be found in Table~\ref{table:algorithms}. 

\begin{table*}
\resizebox{2.1\columnwidth}{!}{\begin{tabular}{|c|c|c|}
\hline
 & VMC \& GFMC & AFMC \\ \hline
Walker state $\phi$ & Basis Slater determinant & General Slater determinant \\ \hline
\begin{tabular}{c}Classical-computing\\trial state\end{tabular} & \begin{tabular}{c}Jastrow-type states (e.g. J-KSzGHF)~\cite{Mahajan2019},\\Coupled cluster~\cite{Roggero2013}, Tensor network~\cite{Sandvik2007} {\it et al.}\end{tabular} & Hartree-Fock, Density-functional theory  {\it et al.}~\cite{motta2018ab} \\ \hline
\begin{tabular}{c}Quantum-computing\\trial state\end{tabular} & \begin{tabular}{c}Unitary coupled cluster,\\Hardware-efficient {\it et al.}\end{tabular} & \begin{tabular}{c}A variant of coupled cluster~\cite{Huggins2022}, Unitary coupled cluster,\\Hardware efficient {\it et al.}\end{tabular} \\ \hline
\begin{tabular}{c}Amplitude-estimation\\quantum circuit\end{tabular} & \begin{tabular}{c}Hadamard test, vacuum reference,\\Hartree-Fock reference\end{tabular} & Hadamard test, vacuum reference \\ \hline
\begin{tabular}{c}Oscillatory-sign-free\\approximation\end{tabular} & fixed-node GFMC & phaseless AFMC \\ \hline
\end{tabular}}
\caption{
Variational Monte Carlo (VMC), Green's function Monte Carlo (VMC) and auxiliary-field Monte Carlo (AFMC) algorithms with a classical-computing trial state or a quantum-computing trial state. Unitary-coupled-cluster and Hardware-efficient trial states can be prepared and optimised using the variational quantum eigensolver algorithm~\cite{cerezo2021variational}. 
}
\label{table:algorithms}
\end{table*}

\section{Consistent and energy evaluation algorithms}
\label{sec:algorithm}

The bias of the ground-state energy in VMC, fn-GFMC and ph-AFMC is determined by the trial state. Conventionally, options of the trial state are limited in each of the algorithms, which must permit the efficient computing of amplitudes $\braket{\Psi_T}{\phi}$ on a classical computer. See Table~\ref{table:algorithms} for examples. In quantum-assisted algorithms, one can choose more general trial states, for example the variant coupled-cluster states used for ph-AFMC in Ref.~\cite{Huggins2022}. In this paper, we generalise this quantum-assisted approach to VMC and fn-GFMC and propose the CQA algorithms. In these algorithms, the QC trial state prepared on a quantum computer is used in two ways: generating walkers $(\phi_l,w_l)$ ($\theta_l = 0$ for these oscillatory-sign-free algorithms) and evaluating the local energy in Eq.~(\ref{eq:EQMC}). As the walker generation is costly, to minimise the QC cost, we further propose QAEE algorithms, in which the quantum computer is only used for evaluating the local energy. 

Before giving the algorithms, we note that all state-related quantities in VMC, GFMC and AFMC can be derived from amplitudes. As introduced previously, the oscillatory-sign-free energy obtained by QMC methods can generally be expressed as  
\begin{align}
E = \frac{\sum_{l = 1}^{N} w_{l} \frac{\bra{\Psi_T}H\ket{\phi_l}}{\braket{\Psi_T}{\phi_l}}}{\sum_{l = 1}^{N} w_{l}}.
\end{align}
In VMC and GFMC, states $\phi=\mathbf{R}$ are basis states, and $\braket{\Psi_T}{\mathbf{R}}$ are amplitudes in the basis. The other quantity we need to compute is the numerator of the local energy, i.e.~$\bra{\Psi_T}H\ket{\mathbf{R}}$. This quantity can be rewritten as $\bra{\Psi_T}H\ket{\mathbf{R}} = \sum_{\mathbf{R}'} \braket{\Psi_T}{\mathbf{R}'}H_{\mathbf{R}',\mathbf{R}}$, where $H_{\mathbf{R}',\mathbf{R}} = \bra{\mathbf{R}'}H\ket{\mathbf{R}}$ are matrix elements of the Hamiltonian. It is required that $H$ is sparse in the basis. Assuming the number of nonzero $H_{\mathbf{R}',\mathbf{R}}$ for the given $\mathbf{R}$ is $L$, we can compute $\bra{\Psi_T}H\ket{\mathbf{R}}$ with $L$ amplitudes. In AFMC, $\phi$ are general SDs instead of basis SDs. In addition to amplitudes, quantities used in AFMC are $\bra{\Psi_T}H\ket{\phi}$ and $\bra{\Psi_T}A\ket{\phi}$, where $A$ are one-particle operators, i.e.~a linear combination of fermion operators $a_p^\dag a_q$ up to a constant term, see Appendix~\ref{app:AFMC}. The Hamiltonian in the second quantisation form is usually formed of one-particle and two-particle terms like $a_p^\dag a_q$ and $a_p^\dag a_{p'}^\dag a_{q'} a_q$. Because $a_p^\dag a_q\ket{\phi}$ and $a_p^\dag a_{p'}^\dag a_{q'} a_q\ket{\phi}$ are also SDs, quantities $\bra{\Psi_T}H\ket{\phi}$ and $\bra{\Psi_T}A\ket{\phi}$ can be expressed as linear combinations of amplitudes in the form $\braket{\Psi_T}{\phi}$. 

This paper reports three sets of algorithms. CQA and QAEE algorithms and their numerical results are presented in this section. In section~\ref{sec:subspace_diag}, we present the QCMCSD algorithm, which is a generic hybrid way to reduce errors in classical algorithms in the framework of quantum-assisted QMC. Sections~\ref{sec:amplitude} and \ref{sec:Bayesian} are subroutines of the algorithms. In Section~\ref{sec:amplitude}, we present three methods to evaluate the amplitude $\braket{\Psi_T}{\phi}$ with a quantum computer. Section~\ref{sec:Bayesian} introduces a Bayesian inference method to reduce the variance due to quantum computing. In section~\ref{sec:symmetry} we discuss the symmetries in fermion systems and propose a method that can reduce the errors based on symmetries. 

\subsection{Consistent quantum-assisted VMC and fn-GFMC}
\label{sec:CQA}

In CQA algorithms, we take a state prepared on the quantum computer as the trial state. For clarity, we use $\ket{\Psi_Q}$ to denote such a QC trial state, and we take $\ket{\Psi_T} = \ket{\Psi_Q}$ in VMC and fn-GFMC (see Algorithms~\ref{alg:VMC} and \ref{alg:fn-GFMC}). However, in quantum computing, we cannot exactly evaluate amplitudes $\braket{\Psi_Q}{\mathbf{R}}$ because of the finite computational resources. Specifically, with a finite number of circuit shots to measure the amplitude, the result has a finite variance. We show that in CQA algorithms there is an effective trial state $\ket{\hat{\Psi}_Q}$ that determines the final result of computation, and the variational property of VMC and fn-GFMC are preserved. We can take advantage of the variational property to reduce the impact of the QC variance. 

Let $\hat{y}_\mathbf{R}$ be the estimate of $\braket{\Psi_Q}{\mathbf{R}}$ obtained in quantum computing. We measure amplitudes in a consistent way as follows: When $\braket{\Psi_Q}{\mathbf{R}}$ is required for the first time, we evaluate it on the quantum computer and record $\hat{y}_\mathbf{R}$ in a classical register; If the same amplitude is required later, instead of reevaluating it, we read the estimate $\hat{y}_\mathbf{R}$ from the register. Compared with reevaluating amplitudes whenever required, the consistent approach reduces the overall QC cost by increasing the CC complexity. 

To introduce the effective trial state, we think of the limit of infinite walkers, i.e.~$N\rightarrow\infty$. In this limit, all relevant amplitudes (irrelevant amplitudes are those forbidden due to certain symmetries) have been evaluated and recorded. Then, the effective trial state reads $\ket{\hat{\Psi}_Q} = \sum_{\mathbf{R}} \hat{y}_\mathbf{R} \ket{\mathbf{R}}$. The bias in VMC and fn-GFMC is determined by this trial state. We remark that in practice the number of walkers is always finite, and only amplitudes queried in the finite walkers are recorded. Notice that if amplitudes are reevaluated every time when it is required, estimates are different every time, and in this case there is not a consistent effective trial state. 

The consistent effective trial state $\ket{\hat{\Psi}_Q}$ can also be constructed using shadow tomography~\cite{Huggins2022}. Our method is proposed in the spirit of importance sampling, i.e.~only those amplitudes queried in VMC or fn-GFMC are evaluated. In this work, we will not benchmark which method is more efficient. We would like to focus on the advantage of using a consistent effective trial state, which is reducing circuit shots according to the variational principle.

In CQA algorithms, the energy is variational, i.e.~higher than the ground-state energy (up to a statistical error due to finite $N$). In VMC, the energy in the $N\rightarrow\infty$ limit is the expected energy of the effective trial state, i.e.~$\frac{\bra{\hat{\Psi}_Q}H\ket{\hat{\Psi}_Q}}{\braket{\hat{\Psi}_Q}{\hat{\Psi}_Q}}$. In fn-GFMC, the fixed-node Hamiltonian is constructed according to the effective trial state, i.e.~replacing $\Psi_T$ with $\hat{\Psi}_Q$ in Appendix~\ref{app:GFMC}, and the corresponding fixed-node energy is always higher than the ground-state energy (using the approach in Refs.~\cite{Blunt2021, Haaf1995}). Therefore, according to the variational principle, we can reduce the bias by repeating the VMC or fn-GFMC calculations a few times and taking the one with the minimum energy as the final result. 

With the variational principle, we can achieve the quantum improvement/advantage at a smaller QC cost. The bias in CQA algorithms is determined by the effective trial state, which is stochastic due to the variance of amplitudes. Effective trial states leading to a bias reduction occurs probabilistically. By post-selecting the lowest energy, a larger QC variance is tolerable in observing the bias reduction. Without the post-selection, we have to reduce the variance to a level that the bias reduction occurs with certainty. With the post-selection, we only need to reduce the variance to the level that bias reduction occurs with a finite probability, e.g.~$50\%$, and the overall probability of bias reduction after the post-selection in a few trials can be high. In Sec.~\ref{sec:Bayesian}, we give a protocol based on Bayesian inference to attain a robust success probability. 

We remark that the effective trial state must be consistent in the entire algorithm. In VMC, we need to sample $\mathbf{R}$ according to the distribution $\abs{\braket{\Psi_Q}{\mathbf{R}}}^2$. This step is straightforward in quantum computing: We can prepare and measure the state $\ket{\Psi_Q}$ in the basis, then the probability of the outcome $\mathbf{R}$ is naturally $\abs{\braket{\Psi_Q}{\mathbf{R}}}^2$. However, to properly implement the consistent approach such that the variational property holds, all quantities must be derived from the same trial state $\ket{\hat{\Psi}_Q}$. Therefore, we need to generate samples with a probability proportional to $\abs{\hat{y}_\mathbf{R}}^2$ instead of $\abs{\braket{\Psi_Q}{\mathbf{R}}}^2$. 

We would like to remark that VMC is discussed here because of its simplicity in concept, such that one can grasp the general idea without getting into details of more complicated algorithms, i.e.~GFMC and AFMC. One may notice that the CQA VMC algorithm is likely not to outperform the VQE algorithm, however, we keep it in this work to demonstrate how the CQA approach works.

\subsection{Quantum-assisted energy evaluation algorithms}

The trial state is used in two phases in QMC algorithms. First, it is used to generate random walkers and yield an expression of the (approximate) ground state shown in Eq.~(\ref{eq:Psi}). Second, given the expression and trial state, the energy in Eq.~(\ref{eq:E}) is evaluated. In Ref.~\cite{Huggins2022} and our CQA algorithms, the quantum trial state is used in both phases. As we see, the walker generation phase is costly. In fn-GFMC and ph-AFMC, the state $\ket{\phi_l}$ of each walker evolves in a stochastic process for many time steps. In each step, we need to evaluate a set of amplitudes $\braket{\Psi_T}{\phi}$. With a QC trial state, i.e.~$\ket{\Psi_T} = \ket{\Psi_Q}$, amplitudes are evaluated on a quantum computer, and a certain number of circuit shots are needed for each amplitude. Note that tens of thousands of walkers and time steps may be required in ph-AFMC~\cite{motta2018ab}, which leads to a large total number of amplitude queries. 

In this section, we propose an alternative way of using the QC trial state, i.e.~QAEE algorithms, to minimize the QC cost. For clarity, we use $\ket{\Psi_Q}$ to denote the QC trial state, which is prepared and measured on the quantum computer, and $\ket{\Psi_C}$ to denote a CC trial state, which can be efficiently evaluated on a classical computer. Some examples of QC and CC trial states of different QMC algorithms can be found in Table~\ref{table:algorithms}. 

In QAEE algorithms, we only use the QC trial state in the energy evaluation phase. First, we take the trial state $\ket{\Psi_T} = \ket{\Psi_C}$ to generate the expression of $\ket{\Psi}$ in Eq.~(\ref{eq:Psi}) entirely on the classical computer. Second, we evaluate the energy in Eq.~(\ref{eq:E}) with the quantum computer, in which we take $\ket{\Psi_T} = \ket{\Psi_Q}$. Therefore, the final energy becomes 
\begin{align}
E = \frac{\sum_{l = 1}^{N} w_{l} \frac{\bra{\Psi_Q}H\ket{\phi_l}}{\braket{\Psi_C}{\phi_l}}}{\sum_{l = 1}^{N} w_{l} \frac{\braket{\Psi_Q}{\phi_l}}{\braket{\Psi_C}{\phi_l}}}.
\label{eq:QAEE}
\end{align}
Here we have taken $\theta_{l} = 0$ in oscillatory-sign-free algorithms. The factor $\frac{\braket{\Psi_Q}{\phi_{l}}}{\braket{\Psi_C}{\phi_{l}}}$ may have a nonzero phase, which can potentially cause the phase problem. However, because both $\ket{\Psi_Q}$ and $\ket{\Psi_C}$ are approximations to the ground state, i.e.~the difference between them is small, we find that the phase problem is negligible in numerical simulations of a $\mathrm{H_4}$ linear chain: in VMC and fn-GFMC the average phases are $0.999995$ and $0.999996$, respectively~\footnote{The average phase is $\langle{e^{i\theta_{Q,l}}}\rangle = (\sum_{l = 1}^{N} w_{Q,l} e^{i\theta_{Q,l}})/(\sum_{l = 1}^{N} w_{Q,l})$, where $w_{Q,l} e^{i\theta_{Q,l}} = w_{l} \frac{\braket{\Psi_Q}{\phi_l}}{\braket{\Psi_C}{\phi_l}}$, $w_{Q,l}$ is positive, and $\theta_{Q,l}$ is real.}. In the case that the problem is severe, we can use the phaseless approximation similar to ph-AFMC to eliminate the phase problem. 

The optimal way to evaluate energy is using the importance sampling method. The denominator of Eq.~(\ref{eq:QAEE}) is a linear combination of amplitudes $\braket{\Psi_Q}{\phi_l}$. Instead of estimating each amplitude with the same QC resources, we assign more circuit shots to an amplitude if the absolute value of its coefficient is larger. We give details of the importance sampling protocol for computing the real part of the denominator using the Hadamard-test circuit (see Sec.~\ref{sec:amplitude}), and it is similar for the imaginary part, numerator and other quantum circuits: 
\begin{enumerate}
    \item Generate $N$ pairs of $\{w_l,\phi_l\}$ on a classical computer with $\ket{\Psi_T} = \ket{\Psi_C}$ according to VMC, fn-GFMC or ph-AFMC; 
    \item Randomly draw a walker $l$ with the probability $P(l)=w_{l}/(C_B\abs{\braket{\Psi_C}{\phi_l}})$, where $C_B$ is the normalisation factor $C_B = \sum_{l = 1}^{N} w_{l}/\abs{\braket{\Psi_C}{\phi_l}}$; 
    \item Compose the Hadamard-test circuit to measure the real part of the amplitude $\braket{\Psi_Q}{\phi_l'}$, where $\ket{\phi_l'} = e^{i\theta'_l}\ket{\phi_l}$, and $\theta'_l = \arg(w_{l}/\braket{\Psi_C}{\phi_l})$. Implement the circuit for one shot to obtain an outcome $\eta = \pm 1$ of the $X$ measurement. Start again from step 2 and repeat $M_{tot}$ times.
\end{enumerate}
Note that in practice, one does not follow steps 2 to 3 for $M_{tot}$ times, but can repeat step 2 for $M_{tot}$ times and record all the walkers $w_l$, before going to step 3 to evaluate the amplitudes on a quantum computer. 
In the above protocol, we have followed the Monte Carlo summation method and rewritten the denominator real part $B = \sum_{l = 1}^{N} w_{l} \Re\left(\frac{\braket{\Psi_Q}{\phi_l}}{\braket{\Psi_C}{\phi_l}}\right)$ in the form of the expected value $B = \sum_{l = 1}^{N} P(l)C_B\Re\left(\braket{\Psi_Q}{\phi_l'}\right)$. For clarity, let $\eta^{(l)}$ be a random variable that represents the outcome of the Hadamard-test circuit for measuring $\Re(\braket{\Psi_Q}{\phi_l'})$, in contrast with $\eta$, which represents the outcome for a randomly chosen $l$. As a property of Hadamard-test circuits, $\eta^{(l)}$ is an unbiased estimator of $\Re(\braket{\Psi_Q}{\phi_l'})$, i.e.~$\mathbf{E}[\eta^{(l)}] = \Re(\braket{\Psi_Q}{\phi_l'})$. Accordingly, $B = C_B\mathbf{E}[\eta]$; let the measurement outcomes be $\{\eta_i| i=1,2,\ldots,M_{tot}\}$, the estimator of $B$ is thus $\hat{B} = (C_B/M_{tot})\sum_{i=1}^{M_{tot}}\eta_i$. As $\eta = \pm 1$, the distribution of $\hat{B}$ is binomial (up to a transformation); and the probability of $\eta = 1$ is $(B/C_B+1)/2$. Therefore, the variance of $\hat{B}$ is $Var[\hat{B}]=(C_B^2-B^2)/M_{tot}\leq C_B^2/M_{tot}$.

Using the QC trial state in the energy evaluation can reduce the error in the energy. Suppose $\{E_i\}$ and $\{\ket{\Phi_i}\}$ are eigenenergies and eigenstates of the Hamiltonian, the energy in Eq.~(\ref{eq:E}) is 
\begin{align}
E = \frac{\sum_i E_i \braket{\Psi_T}{\Phi_i}\braket{\Phi_i}{\Psi}}{\sum_i \braket{\Psi_T}{\Phi_i}\braket{\Phi_i}{\Psi}}.
\end{align}
If $\ket{\Psi_Q}$ is closer to the ground state than $\ket{\Psi_C}$, we expect that the error in $E$ with respect to the ground-state energy is smaller after replacing $\ket{\Psi_T} = \ket{\Psi_C}$ with $\ket{\Psi_T} = \ket{\Psi_Q}$, because magnitudes $\abs{\braket{\Psi_T}{\Phi_i}}$ of excited states are smaller. Besides, in some cases when $\abs{\braket{\Psi_Q}{\Phi_0}}^2$ and $\abs{\braket{\Psi_C}{\Phi_0}}^2$ are close to each other, where $\ket{\Phi_0}$ is the ground state, the final energy from $\ket{\Psi_Q}$ can still be better. This is probably originated from the inherent symmetry projection on the trial state. In Sec.~\ref{sec:symmetry}, we will discuss the symmetries in fermion systems and give such an example. We will show that in quantum-assisted algorithms, the knowledge of symmetry can be used to reduce errors. 

Next, we will demonstrate that the QAEE algorithm can reduce the bias in numerical simulations. Besides the numerical evidence, there are limited theoretical arguments supporting the bias reduction other than the intuitive conjecture: Replacing the trial state in the energy evaluation with a state closer to the true ground state may reduce the bias. This problem will be solved in the QCMCSD algorithm, in which the bias reduction is a rigorous theoretical result under the assumption without statistical error, i.e.~the bias is always smaller than in the classical algorithm. We remark that QCMCSD can also reduce the bias in the quantum algorithm that prepares the QC trial state, such as VQE. In general, QAEE and QCMCSD reduce the requirement on quantum computing for observing quantum advantage, i.e.~we only need to prepare a QC trial state better than the CC trial state to reduce the bias. 

\subsection{Simulation results}
\label{sec:simulation}

We demonstrate quantum-assisted algorithms numerically with the $\mathrm{H_4}$ linear chain. The molecule has four hydrogen atoms with uniform spacing, and the interatomic distance is $0.74 \mathring{A}$. In the STO-3G basis, the number of spin orbitals is eight, which can be encoded into eight qubits using the Jordan-Wigner transformation. We use the J-KSzGHF state~\cite{Mahajan2019} as the CC trial state in VMC and fn-GFMC and the UCCSD state (with one Trotter step~\cite{Barkoutsos2018}) as the QC trial state. Both trial states come with parameters, which are optimized by minimizing the expected energy. For the J-KSzGHF state, the expected energy can be evaluated with the VMC algorithm. For the UCCSD state, the expected energy can be measured on a quantum computer. In our case, we directly compute the energy by exact simulation. Because the molecule is small, both trial states can reach an accurate ground-state energy after optimization. 

The purpose of this numerical simulation is to study the impact of QC variance on the potential quantum advantage. Therefore, we don't intend to maximize the performance of classical algorithms. In each case, we only need a pair of CC and QC trial states, such that without QC variance the bias is smaller if we use the QC trial state. With such two trial states, we can illustrate how the QC variance harms the quantum advantage and to what extent our algorithms can reduce its impact. Therefore, we deliberately stop the optimisation of trial states before the actual minimum is reached. We take the UCCSD trial state with an error of $2.096~\mathrm{mE_h}$ (milli hartree) and the J-KSzGHF trial state with an error of $6.084~\mathrm{mE_h}$ in VMC and $2.068~\mathrm{mE_h}$ in fn-GFMC.

\begin{figure}[tbp]
\begin{center}
\includegraphics[width=1\linewidth]{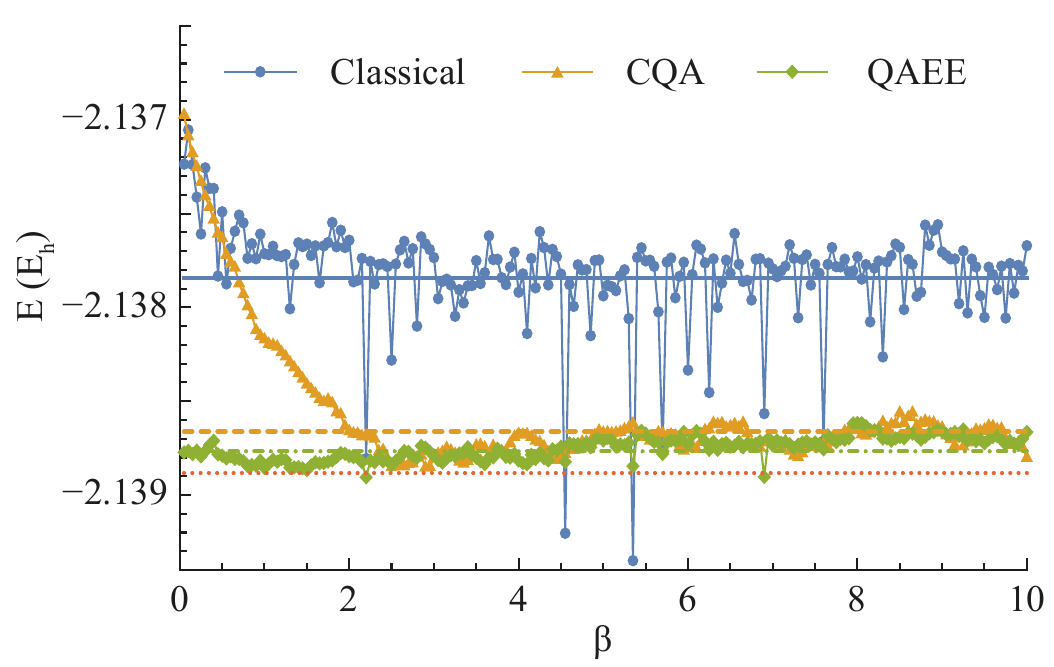}
\caption{
The energy ($\mathrm{E_h}$) of a $\mathrm{H_4}$ linear chain computed using the classical, consistent quantum-assisted (CQA) and quantum-assisted energy evaluation (QAEE) fn-GFMC algorithms. The number of walkers is $N = 10^6$, and $\Delta\beta = 10^{-3}$. The horizontal lines denote energies in the limit $N,\beta\rightarrow\infty$ for the classical (solid), CQA (dashed) and QAEE (dash-dotted) algorithms and the exact ground-state energy (dotted). 
}
\label{fig:GFMC}
\end{center}
\end{figure}

We illustrate the numerical result of the classical and quantum-assisted fn-GFMC with an increasing $\beta$ in Fig.~\ref{fig:GFMC}. A similar result of ph-AFMC can be found in Appendix~\ref{app:afmc:simulation}. We can find that the energy obtained by the classical algorithm has a larger bias and variance than those obtained by quantum-assisted algorithms, even though J-KSzGHF and UCCSD trial states (respectively used in classical and quantum-assisted algorithms) have approximately the same errors in energy. Particularly as $N,\beta\rightarrow\infty$, as shown by the dashed (orange) and dash-dotted (green) curves, the QAEE algorithm leads to an energy closer to the exact ground-state energy than the CQA algorithm.

\begin{table}
\resizebox{\columnwidth}{!}{\begin{tabular}{|c|c|c|}
\hline
 & VMC & fn-GFMC \\ \hline
Classical algorithms & 6.084 & 1.039 \\ \hline
\begin{tabular}{c}CQA ($M\rightarrow\infty$)\end{tabular} & 2.096 & 0.222 \\ \hline
\multirow{2}{*}{\begin{tabular}{c}CQA with\\ Empirical-mean\\estimation\end{tabular}} & \begin{tabular}{c}$M=10000$\\ \hline 10.689 (2.653)\end{tabular} & \begin{tabular}{c}$M=10000$\\ \hline 1.631 (0.657)\end{tabular} \\ \cline{2-3}
& \begin{tabular}{c}$M=50000$\\ \hline 3.843 (0.665)\end{tabular} & \begin{tabular}{c}$M=50000$\\ \hline 0.565 (0.164)\end{tabular} \\ \hline
\multirow{2}{*}{\begin{tabular}{c}CQA with\\ Bayesian inference\\estimation\end{tabular}} &\begin{tabular}{c}$M=10000$\\ \hline 5.198 (0.182)\end{tabular} & \begin{tabular}{c}$M=10000$\\ \hline 0.774 (0.112)\end{tabular} \\ \cline{2-3}
& \begin{tabular}{c}$M=50000$\\ \hline 3.361 (0.226)\end{tabular} & \begin{tabular}{c}$M=50000$\\ \hline 0.383 (0.088)\end{tabular} \\ \hline
\end{tabular}}
\caption{
Errors in the ground-state energy ($\mathrm{mE_h}$) of a $\mathrm{H_4}$ linear chain computed with the classical and consistent quantum-assisted (CQA) algorithms. When circuit shots per amplitude query $M$ is infinite, the energy is calculated with the quantum-computing trial state $\ket{\Psi_T}=\ket{\Psi_Q}$. When $M$ is finite, a Hadamard-test circuit is used when generating the effective trial states, and we calculate the energy with $\ket{\Psi_T}=\ket{\hat{\Psi}_Q}$. In each case, the average energy calculated from 100 effective trial states is given in the table, associated with the corresponding standard deviations shown in the bracket. 
}
\label{table:CQAQMC}
\end{table}

The impact of QC variance in CQA algorithms is summarised in Table~\ref{table:CQAQMC}. Here, we generate the effective trial states with a finite number of circuit shots $M$, then we compute the corresponding expected energy in VMC and fixed-node energy in fn-GFMC, which correspond to the limit $N,\beta\rightarrow\infty$. Two approaches of amplitude estimation are considered. In the first approach, we use the Hadamard-test circuit to directly estimate the amplitude using the empirical mean estimation, see Sec.~\ref{sec:Hadamard}. In the second approach, the Bayesian inference is used, see Sec.~\ref{sec:Bayesian}. For both approaches, we can find that CQA algorithms outperform classical algorithms given a sufficiently large $M$, and the quantum advantage is observed in the Bayesian inference approach when the empirical mean estimation fails.

\begin{table}
\resizebox{\columnwidth}{!}{\begin{tabular}{|c|c|c|}
\hline
 & VMC & fn-GFMC \\ \hline
\begin{tabular}{c}Classical algorithms\\($N=10^6$)\end{tabular} & 6.042 & 1.345 \\ \hline
\multirow{3}{*}{\begin{tabular}{c}QAEE ($N=10^6$) \end{tabular}} & \begin{tabular}{c}$M_{tot}\rightarrow\infty$\\ \hline 0.561\end{tabular} & \begin{tabular}{c}$M_{tot}\rightarrow\infty$\\ \hline 0.180\end{tabular} \\ \cline{2-3}
& \begin{tabular}{c}$M_{tot}=5\times 10^8$\\ \hline 0.542 (0.499)\end{tabular} & \begin{tabular}{c}$M_{tot}=5\times 10^9$\\ \hline 0.163 (0.163)\end{tabular} \\ \hline
\end{tabular}}
\caption{
Errors in the ground-state energy ($\mathrm{mE_h}$) of a $\mathrm{H_4}$ linear chain computed with the classical QMC and quantum-assisted energy evaluation (QAEE) algorithms. We take $N = 10^6$ walkers in Monte Carlo algorithms. When $M_{tot}$ is finite, we compute the energy with a Hadamard-test circuit and the importance sampling protocol; in this case the energy is random because of the variance in quantum computing. In fn-GFMC, we take $\Delta\beta = 10^{-3}$ and the total number of steps $n = 10^4$ (where the energy stops descending and can be regarded as $\beta\rightarrow\infty$), i.e. we get an energy at $\beta=10$. For each case shown in the table, we compute the average of 100 energy samples and the standard deviation is given in the bracket. 
}
\label{table:QAEEQMC}
\end{table}

In the numerical simulation of QAEE algorithms, we take into account the statistical fluctuation due to both finite number of walkers $N$ and number of circuit shots. The results are shown in Table~\ref{table:QAEEQMC}, where we take $N=10^6$ and $\beta=10$ ($\beta$ can be regarded as $\infty$ as the energy stops descending). Besides, an infinite and a finite total number of circuit shots $M_{tot}$ are also considered. As the results indicate, with a finite $N$, the classical algorithm generates an energy close to the $N\rightarrow\infty$ case, see Table~\ref{table:CQAQMC}. Besides, for QAEE algorithms, a finite $M_{tot}$ leads to roughly the same energies as in the case with an infinite $M_{tot}$. 

So far, we have given two ways of using a quantum computer in QMC algorithms, i.e.~CQA algorithms and QAEE algorithms. In QAEE algorithms, the cost of quantum computing is minimized but still can reduce the bias. In general, there are various ways to tune the QC cost. For example, in fn-GFMC, we can take $\ket{\Psi_T} = \ket{\Psi_C}$ and simulate the imaginary-time evolution on a classical computer, then we switch the trial state to $\ket{\Psi_T} = \ket{\Psi_Q}$ and make use of a quantum computer to simulate the evolution for a relatively short time. In this picture, the QAEE algorithm is the extreme case that the quantum computer is only used in the last step.

\section{Quantum computing of amplitudes}
\label{sec:amplitude}

Computing amplitudes $\braket{\Psi_T}{\phi}$ is essential in VMC, GFMC and AFMC. As discussed before, all quantities related to the quantum state used in these QMC algorithms can be derived from amplitudes. There are two approaches to evaluate the amplitude with a quantum computer. In the first approach, one prepares the trial state $\ket{\Psi_T}$ on the quantum computer, implements shadow tomography~\cite{Huang2020} and computes amplitudes on the classical computer using tomography data. This approach has been demonstrated in the experiment reported in Ref.~\cite{Huggins2022}, which has potential scaling issues as pointed out by the authors. The second approach, also mentioned in the same work, is scalable with the system size: Instead of representing the trial state with tomography data, one estimates the specific amplitude $\braket{\Psi_T}{\phi}$ when it is queried. We focus on this non-tomography approach, and techniques developed in this paper can be generalised to the tomography approach. 

In quantum computing, we can accelerate the amplitude estimation using amplitude amplification and quantum phase estimation, in which the error depends on the circuit depth~\cite{Brassard2000}. Because the accelerated estimation usually uses deep circuits, we focus on the direct amplitude estimation, which is more practical for near-term quantum technologies. In this section, we consider three types of quantum circuits for the amplitude estimation, including the conventional Hadamard test~\cite{Ekert2002}, the vacuum-reference protocol~\cite{Lu2021, OBrien2021} and the Hartree-Fock-reference protocol. In the following, we assume that states $\ket{\Psi_T}$ and $\ket{\phi}$ are all normalised. We will show that circuits for VMC and GFMC could be simpler than AFMC. 

\begin{figure}[tbp]
\begin{center}
\includegraphics[width=1\linewidth]{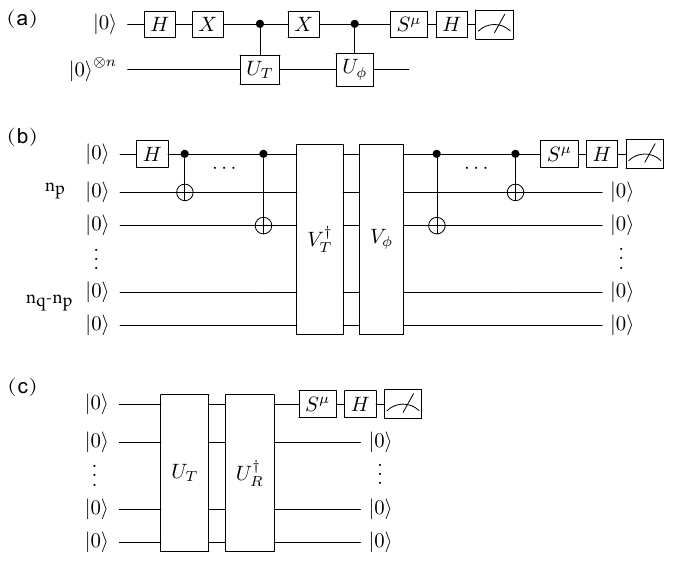}
\caption{
Circuits for amplitude estimation. (a) Hadamard-test circuit.  (b) Vacuum-reference circuit. (c) Hartree-Fock-reference circuit. We take $\mu = 0,3$ for measurements in the $X$ and $Y$ basis, respectively.
}
\label{fig:circuits}
\end{center}
\end{figure}

\subsection{Hadamard test}
\label{sec:Hadamard}

Hadamard test is the conventional method for amplitude estimation~\cite{Ekert2002}. The quantum circuit is shown in Fig.~\ref{fig:circuits}(a), with $n_q$ data qubits and an ancillary qubit. We suppose that unitary operators $U_T$ and $U_\phi$ transform the initial state into the trial state and $\ket{\phi}$, respectively, i.e.~$\ket{\Psi_T} = U_T\ket{0}^{\otimes n_q}$ and $\ket{\phi} = U_\phi\ket{0}^{\otimes n_q}$. The $S_\mu$ gate on the ancillary qubit controls the measurement in the $X$ or $Y$ basis by taking $\mu=0,3$, respectively. The amplitude is $\braket{\Psi_T}{\phi} = \langle{X}\rangle+i\langle{Y}\rangle$. 

To measure the expected value of a Pauli operator $P=X,Y$, we run the corresponding circuit for $M$ shots. Each circuit run has a measurement outcome $\eta_i = \pm 1$. For the empirical mean, the estimator of $\langle{P}\rangle$ is $\frac{1}{M}\sum_{i=1}^M \eta_i$, and the variance is $\frac{1-\langle{P}\rangle^2}{M}$. 

\subsection{Vacuum reference}

We can measure the amplitude without the ancillary qubit in models with the particle number conservation, as proposed in Ref.~\cite{Lu2021, OBrien2021}. We consider the Jordan-Wigner transformation for encoding fermions into qubits, and the number of ones (or zeros depending on the scheme) in the qubit state is the particle number~\cite{ortiz2001quantum}. We suppose that the HF state encoded into qubits is $\ket{HF} = \ket{1}^{\otimes n_p}\otimes\ket{0}^{\otimes(n_q-n_p)}$, where $n_p$ is the particle number. The trial state $\ket{\Psi_T}$ and the walker state $\ket{\phi}$ have the same particle number, then there exist particle-number-preserving unitary operators $V_T$ and $V_\phi$ (i.e.~$[V_T,N_p]=[V_\phi,N_p]=0$, where $N_p$ is the particle number operator) that can transform the HF state into the trial state and $\phi$ state, respectively (i.e.~$\ket{\Psi_T} = V_T\ket{HF}$ and $\ket{\phi} = V_\phi\ket{HF}$). $V_T$ depends on the trial state, and $V_\phi$ is formed of Givens rotation gates~\cite{Kivlichan2018} when $\phi$ is a Slater determinant. 

The circuit is shown in Fig.~\ref{fig:circuits}(b). By measuring the first qubit in $X$ or $Y$ basis and all the other qubits in $\ket{0}$, we effectively measure $\widetilde{X}$ and $\widetilde{Y}$, where $\widetilde{P}=P\otimes\ket{0}\bra{0}^{\otimes n_q-1}$. Then $\braket{\Psi_T}{\phi} = \langle\widetilde{X}\rangle + i\langle\widetilde{Y}\rangle$. The detailed derivation is given in Appendix~\ref{app:vacuum_reference}. This protocol can be used for all three QMC algorithms. 

\subsection{Hartree-Fock reference}

Now, we propose a protocol that only works for VMC and GFMC but uses shallower quantum circuits. In VMC and GFMC, $\ket{\phi} = \ket{\mathbf{R}}$, where $\ket{\mathbf{R}}$ is a basis state. Suppose that the HF state is one of the basis states in $\mathcal{R}$, and each basis state is encoded as a qubit state $\ket{b_1,b_2,\ldots,b_{n_q}}$, where $b_i$ are binary numbers. This condition is satisfied in the Jordan-Wigner transformation~\cite{ortiz2001quantum}. We take the HF state as the reference state, as its overlap with the trial state is usually large. The trial state is prepared by a unitary $\ket{\Psi_T}=U_T\ket{0}^{\otimes n_q}$. 
Without loss of generality, we assume $\braket{\Psi_T}{HF}$ is positive: We can always add a phase factor to $\ket{\Psi_T}$, i.e.~take $\ket{\Psi_T}\leftarrow e^{i\theta}\ket{\Psi_T}$, such that $\braket{\Psi_T}{HF}$ is positive, noticing that the quantum state is the same after changing the phase factor.

Besides, we construct a unitary operator that realises the transformation $\ket{HF} = U_{\mathbf{R}}\ket{0}^{\otimes n_q}$ and $\ket{\mathbf{R}} = U_{\mathbf{R}}\ket{1}\otimes\ket{0}^{\otimes(n_q-1)}$. We suppose that $\ket{HF} = \ket{b_1^{HF},b_2^{HF},\ldots,b_{n_q}^{HF}}$ and $\ket{\mathbf{R}} = \ket{b_1,b_2,\ldots,b_{n_q}}$. Without loss of generality, we suppose that $b_{i_0}^{HF}\neq b_{i_0}$, and $S_{1,i_{0}}$ is a swap gate on the first and $i_0$th qubits. Then, the operator reads 
\begin{align}
U_{\mathbf{R}} = \left(\prod_{i=1}^{n_q} X^{b_i^{HF}}\right) \left(\prod_{i\neq i_0} \Lambda_{i_0,i}^{1-\delta_{b_i,b_i^{HF}}}\right) S_{1,i_{0}},
\end{align}
where $\Lambda_{i_0,i}$ is a controlled-NOT gate with the control qubit $i_0$ and target qubit $i$. We remark that the swap gate is unnecessary in practice, i.e.~we only need to swap operations on the first qubit and qubit $i_0$ when measuring qubits. 

The circuit in Fig.~\ref{fig:circuits}(c) is used to measure the real and imaginary parts of the amplitude $\braket{\Psi_T}{\mathbf{R}}$ by measuring the first qubit (or qubit $i_0$ if the swap gate is removed) in $X$ and $Y$ bases and the rest in state $\ket{0}$, i.e.~effectively measure $\langle\widetilde{X}\rangle$ and $\langle\widetilde{Y}\rangle$, respectively: 
\begin{align}
\braket{\Psi_T}{\mathbf{R}} = \frac{\langle\widetilde{X}\rangle+i\langle\widetilde{Y}\rangle}{2\sqrt{P_r}},
\end{align}
where $P_r=|\braket{\Psi_T}{HF}|^2$. For a detailed derivation, refer to Appendix~\ref{app:hartree_fork}.

Compared with the vacuum reference protocol, the circuit used in the Hartree-Fock reference protocol is usually simpler in two aspects. First, it does not need the GHZ state preparation. Second, the transformation $V_\phi$ used in the vacuum-reference protocol for preparing a general SD can be realised with $O(n_q^2)$ Givens rotation gates. In the Hartree-Fock-reference protocol, the additional transformation $U_{\mathbf{R}}$ is formed of $O(n_q)$ controlled-NOT gates. Therefore, the Hartree-Fock reference protocol has potential advantages in the number of multi-qubit gates.

\section{Bayesian inference amplitude estimation}
\label{sec:Bayesian}

As we have shown, in quantum computing the estimator of $\braket{\Psi_T}{\phi}$ has a finite variance, which could have a significant impact on quantum-assisted QMC algorithms. First, the variance contributes to the bias. For example, in fn-GFMC the bias is the difference between ground-state energies of $H^{fn}$ and $H$~\cite{Blunt2021, Haaf1995}. The sign of the amplitude $\braket{\Psi_T}{\mathbf{R}}$ is crucial for constructing $H^{fn}$. It is difficult to determine the sign if the variance is large in comparison with the absolute value of the amplitude, and CQA fn-GFMC may lose its advantage because of the randomness in signs. Similarly, the phaseless approximation in ph-AFMC depends on the phase of $\braket{\Psi_T}{\phi}$. Second, the amplitude estimation is intensively queried in QMC algorithms, thus reducing the variance with a large number of circuit shots will notably increase the QC cost. 

Here we propose a method for evaluating the amplitude $\braket{\Psi_T}{\phi}$ based on the Bayesian inference. With this method, we can significantly reduce the QC variance (and the measurement cost) and have a stable advantage over the classical algorithm. We consider the trial states $\ket{\Psi_Q}$ and $\ket{\Psi_C}$ as approximations to the ground state. Therefore, $\ket{\Psi_C}$ contains information about $\ket{\Psi_Q}$. Given $\braket{\Psi_C}{\phi}$, we have some prior knowledge about $\braket{\Psi_Q}{\phi}$ before evaluating it on a quantum computer, and we can make use of such information to reduce the QC variance. 

As different quantum circuits are used to measure the real and imaginary parts of the amplitude, the real and imaginary parts are independent and thus can be estimated separately with the Bayesian inference method. In the following, we focus on the real part, and it is the same for the imaginary part. To simplify the expressions, we introduce notations $y_0 \equiv \Re(\braket{\Psi_C}{\phi})$, $y \equiv \Re(\braket{\Psi_Q}{\phi})$ and $x$, where $x$ is the measurement outcome of $\Re(\braket{\Psi_Q}{\phi})$ obtained directly on a quantum computer (e.g.~with Hadamard-test circuits for $M$ shots). In the spirit of the Bayesian method, both $x$ and $y$ are treated as values of random variables, denoted by $\mathbf{X}$ and $\mathbf{Y}$, respectively. To carry out the Bayesian method, we need a prior distribution of $\mathbf{Y}$, which is determined by $y_0$. 

Given a prior distribution $p_{\mathbf{Y}}(y)$, the posterior distribution is 
\begin{align}
p_{\mathbf{Y}\vert \mathbf{X}}(y\vert x) = \frac{p_{\mathbf{X}\vert \mathbf{Y}}(x\vert y)p_{\mathbf{Y}}(y)}{p_{\mathbf{X}}(x)}.
\end{align}
For many methods of measuring $\Re(\braket{\Psi_Q}{\phi})$ on a quantum computer, the distribution $\mathbf{X}\vert \mathbf{Y}$ only depends on $y$. Considering the standard Hadamard-test circuit, $p_{\mathbf{X}\vert \mathbf{Y}}(x\vert y)$ is essentially a binomial distribution, i.e.~$p_{\mathbf{X}\vert \mathbf{Y}}(x\vert y) = {{M}\choose{k}}q^k(1-q)^{M-k}$, where $q=(1-y)/2$, $k = M(1-x)/2$, and $M$ is the number of circuit shots.

The eventual performance of the Bayesian inference depends on the prior. To demonstrate our method, we assume that the prior is a normal distribution centered at $y_0$ with the standard deviation $\sigma_0$: $\mathbf{Y}\sim \mathcal{N}(y_0,\sigma_0)$. We approximate the binomial distribution $p_{\mathbf{X}\vert \mathbf{Y}}(x\vert y)$ with the normal distribution (which can be done when $M$ is large), i.e.~$\mathbf{X}\vert \mathbf{Y}\sim \mathcal{N}(y,\sigma_Q)$, where $\sigma_Q = \sqrt{\frac{1-y^2}{M}}$. For simplification, we can further approximate $\sigma_Q$ with $\sqrt{\frac{1-y_0^2}{M}}$ or its upper bound $\sqrt{\frac{1}{M}}$. Under these approximations (specifically, we take the upper bound in the following), the standard deviation of $\mathbf{X}\vert \mathbf{Y}$ is independent of $y$, and the posterior distribution $\mathbf{Y}\vert \mathbf{X}$ is normal, i.e.~$\mathbf{Y}\vert \mathbf{X}\sim \mathcal{N}(\hat{y},\hat{\sigma})$, where 
\begin{align}
\hat{y} = \frac{y_0+\sigma_0^2Mx}{1+\sigma_0^2M},
\label{eq:haty}
\end{align}
and 
\begin{align}
\hat{\sigma} = \frac{\sigma_0}{\sqrt{1+\sigma_0^2M}}.
\end{align}
Note that $x\rightarrow y$ when the number of samples $M \rightarrow\infty$. We take $\hat{y}$ as the final estimate of $y$. 

The benefit of the Bayesian inference method comes from two sides. On the one hand, quantum computing provides a correction to the classical amplitude $y_0$. The correction is more evident when the quantum computing result is more certain, i.e.~$M$ is larger. Eventually, in the limit $M\rightarrow\infty$, the final estimate $\hat{y}$ converges to its true value. On the other hand, when $x$ deviates from $y$, e.g.~in the case that $M$ is small, the classical amplitude $y_0$ can instead serve as a correction to the quantum estimate to stabilise the final estimate $\hat{y}$. Consequently, even if $M$ is small, we expect that quantum-assisted algorithms can still outperform classical algorithms up to some controllable fluctuation. 

\begin{figure}[tbp]
\begin{center}
\includegraphics[width=1\linewidth]{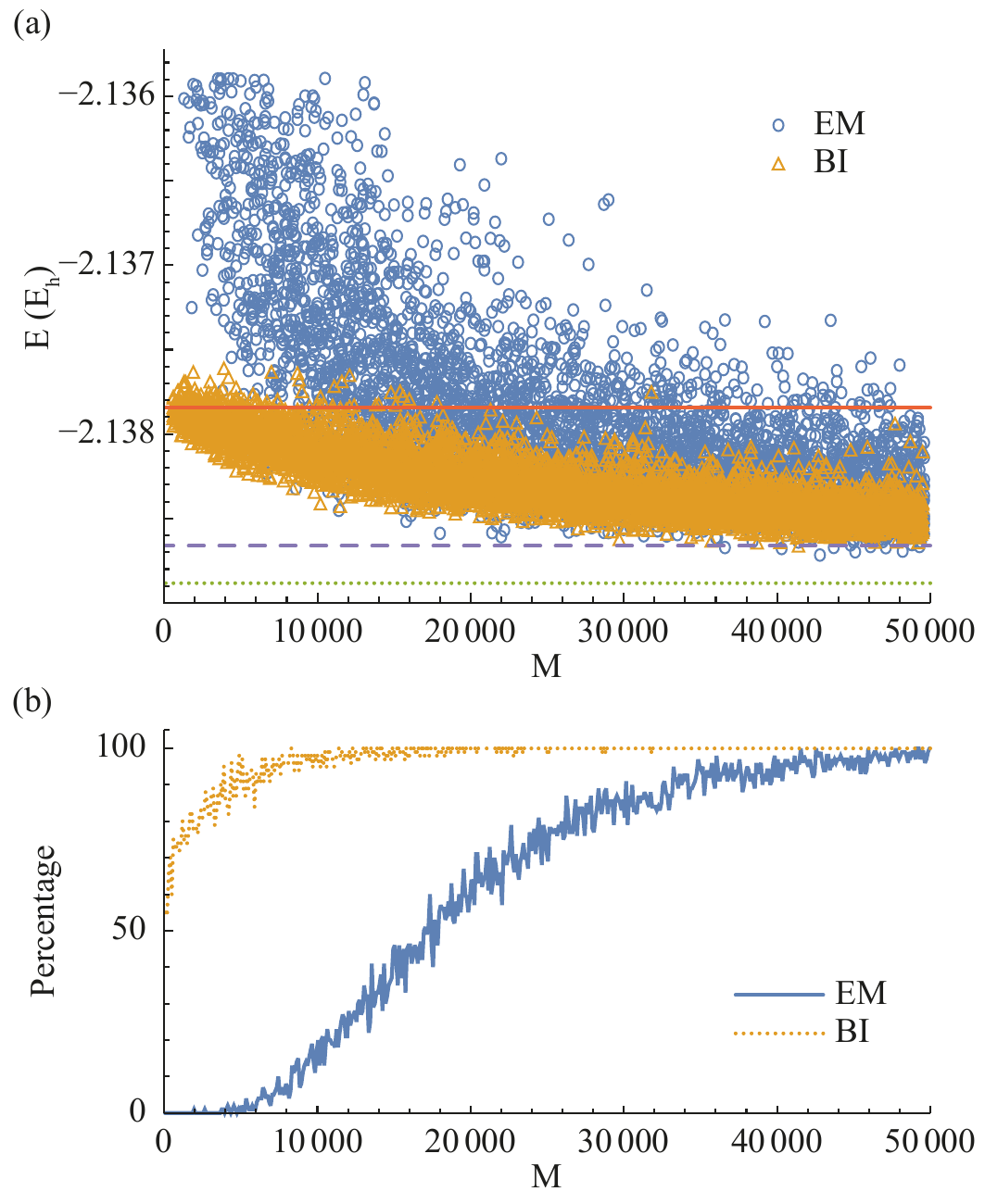}
\caption{
(a) The energy ($\mathrm{mE_h}$) of a $\mathrm{H_4}$ linear chain computed using the consistent quantum-assisted fn-GFMC. $M$ is the number of circuit shots for evaluating an amplitude $\braket{\Psi_Q}{\mathbf{R}}$. We generate effective trial states with the empirical mean (EM) estimation and Bayesian inference (BI) estimation for various $M$. A hundred effective trial states are generated for each $M$, and fixed-node energies of ten states are plotted in the figure. In the BI estimation, we take $\sigma_0^2 = 10^{-5}$. The horizontal lines denote the fixed-node energies of the convectional J-KSzGHF state (solid), quantum-computing UCCSD state (dashed), and the true ground-state energy (dotted). (b) The percentage of samples below the fixed-node energy of the J-KSzGHF state in these 100 samples. 
}
\label{fig:GFMC_BI}
\end{center}
\end{figure}

Numerical simulations are conducted to compare the Bayesian inference estimation [i.e., compute $\hat{y}$ according to Eq.~(\ref{eq:haty})] with the empirical mean estimation (i.e., directly take $\hat{y} = x$) when evaluating the amplitude. We implement the CQA fn-GFMC algorithm here as the example, with parameter settings the same as in Table~\ref{table:CQAQMC}. The results are shown in Fig.~\ref{fig:GFMC_BI}. We find that the energy from the Bayesian inference method is much more stable than that from the empirical-mean method. When $M$ is small (i.e.~the QC variance is large), the quantum-assisted algorithm using the empirical-mean method has a larger bias than the classical algorithm; using the Bayesian inference method, the bias is smaller. When $M$ is large, the energy converges faster to the limit set by the exact QC trial state (i.e.~$M\rightarrow \infty$) using the Bayesian inference method. Results of VMC are similar as summarised in Table~\ref{table:CQAQMC}. 

Using the Bayesian inference method, we can achieve the quantum advantage, i.e.~a reduced bias, with a relatively small $M$. When $M = 100$ (the minimum number taken in the simulation), the bias in the CQA fn-GFMC is smaller than the classical algorithm with a probability of about $50\%$. When $M = 3200$, the bias is reduced by $10\%$ with a probability of $50\%$. Notice that the energy is variational, and we can repeat the computation to generate a set of energies and choose the lowest energy, as discussed in Sec.~\ref{sec:CQA}. Then, the $50\%$ success probability means that we can achieve the quantum advantage by repeating the computation two times on average. 

Using $\braket{\Psi_C}{\phi}$ as the prior guess of $\braket{\Psi_Q}{\phi}$, we have assumed that $\ket{\Psi_C}$ is approximately normalised, as the state $\ket{\Psi_Q}$ prepared on the quantum computer is normalised. In AFMC, we can take $\ket{\Psi_C}$ as a normalised SD (e.g.~HF state). In VMC and GFMC, we can construct $\ket{\Psi_C}$ by applying a Jastrow factor to a normalised SD. However, the normalisation assumption generally does not always hold, e.g.~when $\ket{\Psi_C}$ is a J-KSzGHF state. In this case, we can process the state as follows. We measure $\braket{\Psi_Q}{\phi}$ for a selected state $\phi$, for instance the HF state, and compute $\braket{\Psi_C}{\phi}$. Then, we multiply $\ket{\Psi_C}$ by a factor and take $\ket{\Psi_C'} = \frac{\braket{\Psi_Q}{\phi}}{\braket{\Psi_C}{\phi}}\ket{\Psi_C}$. Then, we use $\ket{\Psi_C'}$ as the prior instead of $\ket{\Psi_C}$ in the Bayesian inference method. 

\section{Symmetry projected quantum-computing trial state}
\label{sec:symmetry}

The ground state is usually in a subspace due to the symmetries of the Hamiltonian. Applying a projection operator of the subspace on a state removes its components orthogonal to the subspace, which is irrelevant to the ground state. In general the projection improves the approximation of a trial state to the ground state. In classical computing, we can introduce correlations to mean-field states by applying the symmetry projection, which is used when generating the trial states in VMC and fn-GFMC, e.g.~KSzGHF states~\cite{Mahajan2019}. In this paper, we propose that in (fully or partially) quantum-assisted QMC algorithms, we can project the QC trial state according to certain symmetries in order to reduce errors. 

Verifying the symmetry in quantum computing can be used to reduce error, and it is a commonly-used method in error mitigation~\cite{endo2022quantum,cai2021quantum}; however, such a technique usually increases the circuit complexity. Specifically, in a quantum computer, we can directly measure qubits in the computation basis, i.e.~the basis of Pauli $Z$ operators. To verify the symmetry in quantum computing, we need to transform the symmetry (i.e.~the corresponding observable) to an observable that can be directly measured. Such a transformation is realised using quantum gates, which increases the circuit complexity~\cite{sack2022avoiding,grant2019initialization}. Depending on the measurement outcome, the state is probabilistically projected onto the subspace or a state orthogonal to the subspace. On the other hand, in quantum-assisted QMC algorithms, we can apply the symmetry projection without additional gates.

\begin{theorem}
Let $P$ be the orthogonal projection onto a subspace $\mathcal{H}_S$ of the Hilbert space, and $[P,H] = 0$. If all walker states $\ket{\phi}$ are in $\mathcal{H}_S$, the energy in Eq.~(\ref{eq:E}) computed with the trial state $\ket{\Psi_T} = \ket{\Psi_Q}$ is the same as the energy computed with $\ket{\Psi_T} = P\ket{\Psi_Q}$. 
\label{the1}
\end{theorem}

The proof is straightforward. Because $\ket{\phi}\in \mathcal{H}_S$ for all $\phi$, $P\ket{\Psi} = \ket{\Psi}$. Then 
\begin{align}
\frac{\bra{\Psi_Q}H\ket{\Psi}}{\braket{\Psi_Q}{\Psi}} = \frac{\bra{\Psi_Q}HP\ket{\Psi}}{\bra{\Psi_Q}P\ket{\Psi}} = \frac{\bra{\Psi_Q}PH\ket{\Psi}}{\bra{\Psi_Q}P\ket{\Psi}}.
\end{align}
According to Theorem~\ref{the1}, the QC trial state is effectively projected if we restrict walker states in the subspace. In the following, we show this restriction on walker states is for free in VMC, GFMC and AFMC. 

For fermion systems, the symmetries usually include the particle number $N_p$, spin component $S_z$, etc. In VMC and GFMC, it is natural that we choose the basis according to these symmetries. We take the $\mathrm{H_2}$ molecule as an example. In the STO-3G basis, the model has four spin orbitals filled with two electrons. We use $A\uparrow$, $A\downarrow$, $B\uparrow$ and $B\downarrow$ to label the four spin orbitals, in which $A,B$ denote two orbitals, and $\uparrow,\downarrow$ denote spins, respectively. The ground state has the $S_z$ symmetry, and the component of the total spin in the z direction is zero. Therefore, one electron is in $A\uparrow$ and $B\uparrow$, and the other electron is in $A\downarrow$ and $B\downarrow$. The dimension of the entire Hilbert space of four spin orbitals is $16$, and the subspace dimension with $N_p$ and $S_z$ symmetries is four. The four states are $\ket{1_{A\uparrow},0_{B\uparrow},1_{A\downarrow},0_{B\downarrow}}$, $\ket{1_{A\uparrow},0_{B\uparrow},0_{A\downarrow},1_{B\downarrow}}$, $\ket{0_{A\uparrow},1_{B\uparrow},1_{A\downarrow},0_{B\downarrow}}$ and $\ket{0_{A\uparrow},1_{B\uparrow},0_{A\downarrow},1_{B\downarrow}}$, which respect the two symmetries and are sufficient for expressing the ground state. Therefore, in VMC and GFMC, we can take $\ket{\mathbf{R}}$ only in these four states. In AFMC, using the Hubbard-Stratonovich transformation, we express the imaginary-time propagator as an integral of one-particle propagators, and we can let the one-particle propagators preserve the $S_z$ symmetry of the Hamiltonian, i.e.~$A$ operators (see Appendix~\ref{app:AFMC}) have the symmetry between two spins. With $\ket{1_{A\uparrow},0_{B\uparrow},1_{A\downarrow},0_{B\downarrow}}$ as the initial state (we suppose that it is the HF state), the initial state is in the four-dimensional subspace, then the walker state is always in the subspace because of the symmetry of $A$ operators, i.e.~walker states $\phi_l$ are SDs with the $N_p$ and $S_z$ symmetries. 

We can generalise this approach to symmetries other than $N_p$ and $S_z$. For example, for the $\mathrm{H_2}$ molecule, we can take $\ket{\mathbf{R}}$ only from the three states $\ket{1_{A\uparrow},0_{B\uparrow},1_{A\downarrow},0_{B\downarrow}}$, $\ket{0_{A\uparrow},1_{B\uparrow},0_{A\downarrow},1_{B\downarrow}}$ and $\frac{1}{\sqrt{2}}(\ket{1_{A\uparrow},0_{B\uparrow},0_{A\downarrow},1_{B\downarrow}}+\ket{0_{A\uparrow},1_{B\uparrow},1_{A\downarrow},0_{B\downarrow}})$, such that the $\ket{\Psi_Q}$ is effectively projected onto the subspace with the $N_p$, $S_z$ and $S^2$ (total spin) symmetries. In many models, e.g.~molecules without an external magnetic field, the Hamiltonian is real, and the ground state is real, which is called the complex conjugation symmetry. If $\ket{\Psi}$ is real, such as in VMC and fn-GFMC, we can impose the complex conjugation symmetry by modifying Eq.~(\ref{eq:E}) to 
\begin{align}
E = \frac{\Re\left(\bra{\Psi_Q}\right)H\ket{\Psi}}{\Re\left(\bra{\Psi_Q}\right)\ket{\Psi}} = \frac{\Re\left(\bra{\Psi_Q}H\ket{\Psi}\right)}{\Re\left(\braket{\Psi_Q}{\Psi}\right)}.
\end{align}

\begin{figure}[tbp]
\begin{center}
\includegraphics[width=1\linewidth]{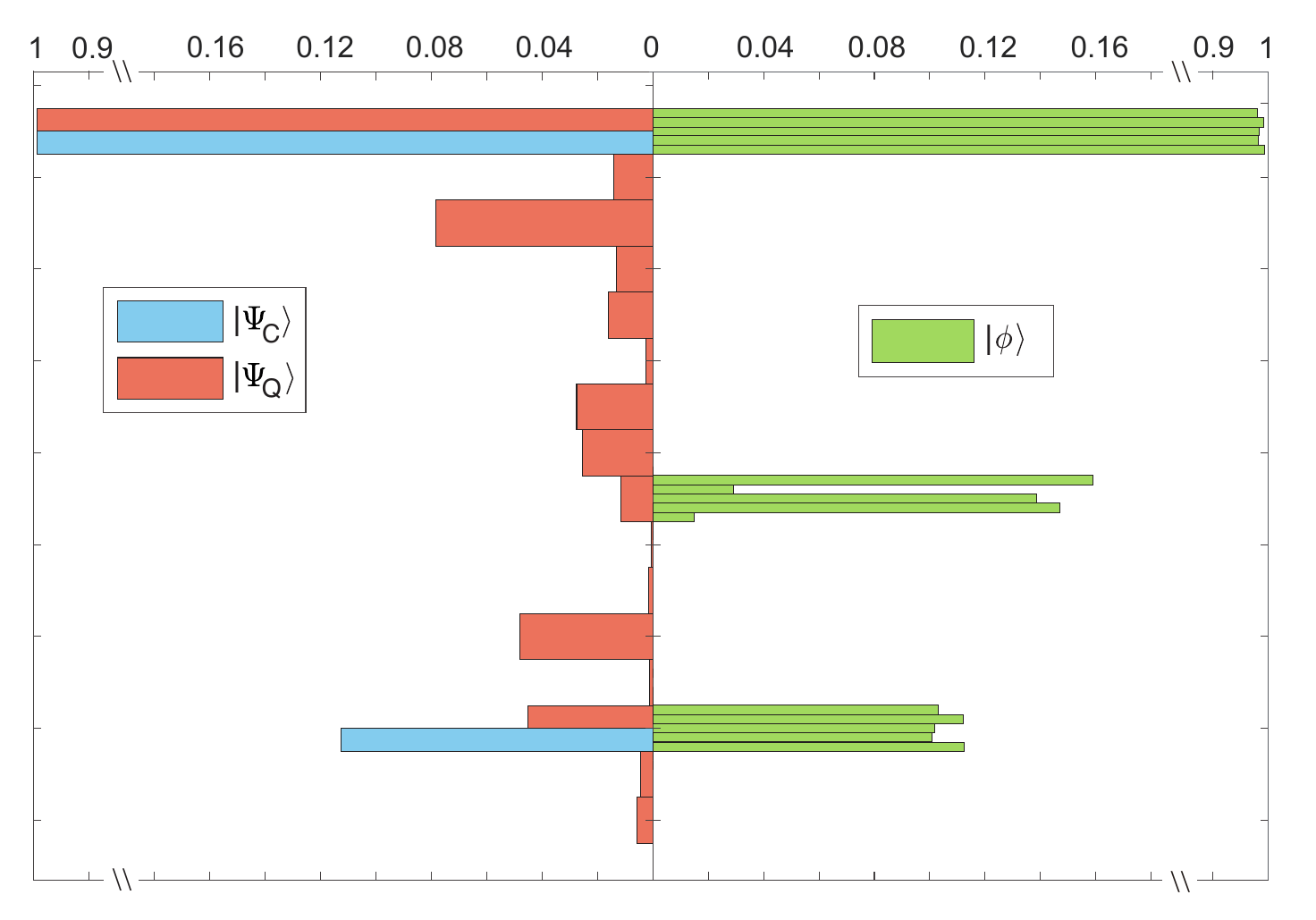}
\caption{
The absolute value of the amplitude of quantum-computing trial state, classical-computing trial state and walker states in the eigenbasis of the $\mathrm{H_2}$ Hamiltonian. 
}
\label{fig:ampli_basis}
\end{center}
\end{figure}

The inherent symmetry projection effectively improves the quality of the QC trial state. It turns out that even in the case CC and QC trial states have roughly the same error in energy (or have approximately the same state fidelity with respect to the ground state), the latter can lead to a smaller error in the final ground state energy, see Fig.~\ref{fig:GFMC} (and Appendix~\ref{app:afmc:simulation}) for the numerical example. We remark that symmetry may not be the only reason for this phenomenon. To illustrate the impact of symmetry projection, we decompose CC and QC trial states with approximately the same state fidelity (which are used in the simulation presented in Appendix~\ref{app:afmc:simulation}) into eigenstates $\ket{\Phi_i}$ of the Hamiltonian of a $\mathrm{H_2}$ molecule and plot amplitudes in Fig.~\ref{fig:ampli_basis}. For simplification and better illustration, we plot the absolute value of the amplitude $|\braket{\Psi_T}{\Phi_i}|$. From the left side, we see that the CC trial state has non-zero components only on two eigenstates, while the QC trial state has a more uniform distribution except for the ground state (the top bar). The right side shows amplitudes $|\braket{\phi}{\Phi_i}|$ of five walker states which we picked randomly in the simulations. We find walker states have non-zero components only on three eigenstates due to symmetries. Because of the symmetries of walker states, the QC trial state is effectively projected onto the same three eigenstates, and components on other eigenstates are effectively removed. After the projection, the fidelity of the QC trial state is increased. This may explain why with a similar state fidelity the QC trial state can lead to a smaller error in the ground-state energy. 

\section{Quantum-classical Monte Carlo subspace diagonalisation}
\label{sec:subspace_diag}

In CQA algorithms, the energy as the result is variational. Because of this reason, the quantum advantage, i.e.~a reduced bias, is verifiable: If the energy is lower than those from classical VMC and fn-GFMC algorithms, the bias of quantum-assisted algorithms is smaller. However, ph-AFMC and QAEE algorithms do not have this property, and it could be the case that the energy is even lower than the true ground-state energy; in this case a lower energy may not be better. Therefore, we would like to ask the question, is there a way to reduce the bias that is systematic, verifiable and generalisable to other advanced classical algorithms?

Diagonalisation in a subspace is one of the most important methods for computing extreme eigenvalues. In classical computing, it is standard to choose the Krylov subspace generated by applying the Hamiltonian power on a reference state~\cite{ozaki2006n}. This approach has been generalised to quantum computing, in which a set of states prepared on the quantum computer spans the subspace~\cite{nakanishi2019subspace,seki2021quantum,cortes2022quantum}. A crucial property of subspace diagonalisation is that, according to the Rayleigh quotient theorem, the lowest eigenenergy is the lowest expected energy in the subspace. In other words, given the best solution to the ground state from classical computing, if we construct a subspace containing this classical solution and an optimal state from quantum computing, the diagonalisation in this subspace always yields an energy closer to the true ground-state energy. This property is exactly what we need. 

To carry out the subspace diagonalisation involving optimal classical and quantum solutions, we need to prepare the classical-solution state on the quantum computer. For applications on near-term quantum devices, it is essential to reduce the gate number. In the following, we show that using the hybrid QMC method the state preparation requires $O(n_q)$ gates, where $n_q$ is the qubit number. 

In VMC, GFMC and AFMC algorithms, we express the approximation to the ground state $\ket{\Psi}$ as a linear combination of walker states $\ket{\phi}$, see Eq.~(\ref{eq:Psi}). As we have discussed in this paper, we can evaluate quantities $\braket{\Psi_Q}{\Psi}$ and $\bra{\Psi_Q}H\ket{\Psi}$ without physically preparing the state $\ket{\Psi}$, i.e.~we only need to estimate amplitudes $\braket{\Psi_Q}{\phi}$. This feature is particularly useful for problems where $\ket{\Psi}$ is harder to prepare than $\ket{\phi}$. Such states include solutions of state-of-the-art methods, for instance coupled cluster~\cite{Roggero2013} and tensor network~\cite{shi2006classical,Sandvik2007}, which can be prepared on a quantum computer: Preparing a coupled-cluster state requires $O(n^4)$ gates~\cite{xia2020qubit}, and it is similar for a tensor-network state~\cite{felser2021efficient}. In VMC and fn-GFMC, to express $\ket{\Psi}$ as a linear combination of basis states to estimate $\braket{\Psi_Q}{\Psi}$ [the denominator of Eq.~(\ref{eq:QAEE})], preparing $\phi$ only requires at most $n_q$ single-qubit gates when $\phi$ is a basis state. Therefore, in the framework of QMC, the circuit complexity can be reduced. It is similar for $\bra{\Psi_Q}H\ket{\Psi}$. 

In the hybrid Monte Carlo subspace diagonalisation, we construct the subspace as follows. Let $\ket{\Psi_C}$ and $\ket{\Psi_Q}$ be optimal solutions from classical and quantum computing, respectively. We can use VQE algorithms to find a quantum solution. The classical solution is from, for instance, QMC, coupled-cluster or tensor-network algorithms. For example, thinking of VMC, we have found the optimal parameters $\boldsymbol{\lambda}^{\star}$, and the corresponding final solution is $\ket{\Psi_C} = \ket{\Psi_T(\boldsymbol{\lambda}^{\star})}$; Together with the exact expression, VMC also yields a walker-state representation of $\ket{\Psi_T(\boldsymbol{\lambda}^{\star})}$, see Eq.~(\ref{eq:PsiVMC}). In GFMC, the state closest to the true ground state is $e^{-n\Delta\beta H}\ket{\Psi_I}$. Instead of an exact expression of this state, fn-GFMC outputs a walker-state representation $\ket{\Psi}$ as an approximation to $e^{-n\Delta\beta H}\ket{\Psi_I}$: $\phi_l$ and $w_l$ in Eq.~(\ref{eq:Psi}) are generated according to Algorithm~\ref{alg:fn-GFMC}, where $\theta_l = 0$. It is similar for ph-AFMC. For coupled-cluster and tensor-network states, we can also work out the corresponding walker-state representation following Refs.~\cite{Roggero2013, Sandvik2007}. In general, any classical-solution state with amplitudes in a basis can be efficiently calculated, and we can work out a walker-state representation according to VMC. Overall, we suppose that there is a walker-state representation $\ket{\Psi}$ of $\ket{\Psi_C}$ in the form of Eq.~(\ref{eq:Psi}). The subspace is the span of two states $\ket{\Psi}$ and $\ket{\Psi_Q}$. This approach can be directly generalised to multiple classical and quantum solutions. 

The subspace diagonalisation works as follows. Let $\ket{\Phi_1} = \ket{\Psi}$ and $\ket{\Phi_2} = \ket{\Psi_Q}$. We need to work out the matrix elements $\mathbf{H}_{i,j} = \bra{\Phi_i}H\ket{\Phi_j}$ and $\mathbf{S}_{i,j} = \braket{\Phi_i}{\Phi_j}$, where $i,j=1,2$. Elements $\mathbf{H}_{1,1}$ and $\mathbf{S}_{1,1}$ can be calculated on the classical computer. Because $\ket{\Psi_Q}$ is physically prepared on the quantum computer, it is normalized, and $\mathbf{S}_{2,2}=1$; $\mathbf{H}_{2,2}$ can be evaluated on the quantum computer. For off-diagonal elements, they can be evaluated as the same as $\bra{\Psi_Q}H\ket{\Psi}$ and $\braket{\Psi_Q}{\Psi}$ in QAEE algorithms. Given two matrices, we solve the generalised eigenvalue problem $\mathbf{H}\mathbf{x} = E\mathbf{S}\mathbf{x}$, where $\mathbf{x}$ is a column vector~\cite{wall1995extraction,epperly2021theory}. The minimum eigenvalue $E$ is the lowest expected energy of all states (including $\ket{\Psi_C}$ and $\ket{\Psi_Q}$) in the subspace.

\begin{table}
\resizebox{0.9\columnwidth}{!}{\begin{tabular}{|c|c|c|}
\hline
 & VMC & fn-GFMC \\ \hline
\begin{tabular}{c}Classical algorithms\\($N=10^6$)\end{tabular} & 6.042 & 1.345 \\ \hline
\begin{tabular}{c}Subspace\\diagonalisation\end{tabular} & 1.761 & 0.861 \\ \hline
\end{tabular}}
\caption{
Errors in the ground-state energy ($\mathrm{mE_h}$) of a $\mathrm{H_4}$ linear chain computed with the classical Monte Carlo and quantum-classical Monte Carlo subspace diagonalisation algorithms. In the classical algorithms, we take $N = 10^6$ and the same trial states as in Table~\ref{table:QAEEQMC}.
}
\label{table:QCMCSD}
\end{table}

To demonstrate the QCMCSD algorithm, we consider VMC and fn-GFMC for classical solutions and VQE with UCCSD for the quantum solution, see Sec.~\ref{sec:simulation}. We remark that the optimisation in VMC and VQE can be further pushed to produce better solutions than what we use in the numerical simulation, and these pseudo-optimal solutions are sufficient for the preliminary demonstration. The result is summarised in Table~\ref{table:QCMCSD}. Expressed as a linear combination of $10^6$ walkers according to VMC, the energy of $\ket{\Psi}$ has an error of $6.042~\mathrm{mE_h}$. $\ket{\Psi_Q}$ is a UCCSD state with an error of $2.096~\mathrm{mE_h}$. The diagonalisation in the subspace $\mathrm{Span}(\{\ket{\Psi},\ket{\Psi_Q}\})$ results in a minimum eigenvalue with the error of $1.761~\mathrm{mE_h}$. The result is similar for fn-GFMC, where we take $\ket{\Psi}$ as a walker-state representation of the fixed-node ground state. This approach can also be applied to the output $\ket{\Psi}$ of ph-AFMC. 

In the numerical result, we find that the QCMCSD algorithm is less accurate than other quantum-assisted algorithms. However, it is of particular interest. As far as we know, there is no rigorous theory showing that utilising a QC trial state can reduce the bias with certainty, although the bias reduction is observed in experiments~\cite{Huggins2022} and numerics. According to the theory of subspace diagonalisation, the QCMCSD algorithm always results in a smaller bias than the classical algorithm, under the assumption of no statistical error.

\section{Conclusion}
\label{sec:conclusion}

In summary, this work introduces a set of quantum-assisted Monte Carlo algorithms. Combining quantum computing with QMC, the advanced classical computational techniques, one can reduce the bias when constraining the sign problem in computing the ground-state energy. This form of quantum advantage is achieved by utilising a trial state prepared on the quantum computer. The potential challenge for hybrid Monte Carlo algorithms of this kind is the cost for measuring trial state amplitudes. We propose these algorithms with the purpose to 
ease this issue. Our methodology is to control the involvement of the quantum trial state, specifically in two ways. First, we use both the quantum trial state and a trial state of classical computing in different stages of the computation. We use the quantum trial state in the entire computation in CQA algorithms and only at the last stage in QAEE algorithms, and intermediate cases are possible. Second, we take the quantum trial state as an adaptable correction (depending on the cost budget) to the classical trial state in the Bayesian inference amplitude estimation. Additionally, applying post-selections in CQA algorithms lowers the requirement for the precision of amplitude measurement. In this work, we numerically demonstrate the performance of our algorithms with hydrogen molecules. Overall, theoretical and numerical results suggest that we can avoid the high-precision amplitude measurement to attain a bias reduction. 

The hybrid Monte Carlo framework provides new tools for quantum computing. We can reduce errors in a quantum trial state with the inherent symmetry projection, which is almost free in Monte Carlo algorithms. We conjecture that this symmetry projection can also mitigate machine noise in quantum computing. The hybrid Monte Carlo is a gate-efficient interface between quantum computing and classical computing to leverage both computation paradigms. In this paper, we propose the QCMCSD algorithm as an example. QMC is a numerical method developed for decades and encompasses a family of variants. In this view, hybrid Monte Carlo offers a powerful suite of techniques to explore to demonstrate practical quantum advantage. 

\begin{acknowledgments}
We acknowledge the support of the National Natural Science Foundation of China (Grants No. 11875050 and No. 12088101) and NSAF (Grant No. U1930403). 
\end{acknowledgments}

\onecolumn\newpage

\bibliographystyle{unsrtnat}
\bibliography{reference}

\onecolumn\newpage
\appendix

\section{Green's function Monte Carlo}
\label{app:GFMC}

\begin{figure*}
\begin{minipage}{\linewidth}
\begin{algorithm}[H]
{\small
\begin{algorithmic}[1]
\caption{{\small Energy evaluation in the variational Monte Carlo algorithm.}}
\label{alg:VMC}
\Statex
\State Input $H$, $\Psi_T$, $N$. 
\For{$l=1$ to $N$}
\State Generate $\mathbf{R}$ according to the probability distribution $\abs{\braket{\Psi_T}{\mathbf{R}}}^2$. 
\State Compute the local energy $E_l = E^{\mathrm{loc}}(\mathbf{R})$ according to Eq.~(\ref{eq:Eloc}). 
\EndFor
\State Output the energy $E\leftarrow \frac{1}{N}\sum_{l=1}^N E_l$. 
\end{algorithmic}
}
\end{algorithm}
\end{minipage}
\end{figure*}

\begin{figure*}
\begin{minipage}{\linewidth}
\begin{algorithm}[H]
{\small
\begin{algorithmic}[1]
\caption{{\small Fixed-node Green's function Monte Carlo algorithm.}}
\label{alg:fn-GFMC}
\Statex
\State Input $H$, $\Psi_I$, $\Psi_T$, $N$, $n$, $\Delta \beta$. 
\For{$l=1$ to $N$}
\State $A \leftarrow \sum_{\mathbf{R}''} \braket{\Psi_T}{\mathbf{R}''}\braket{\mathbf{R}''}{\Psi_I}$
\State Generate $\mathbf{R}$ according to the probability distribution $A^{-1}\braket{\Psi_T}{\mathbf{R}}\braket{\mathbf{R}}{\Psi_I}$. 
\State $w \leftarrow A$
\For{$k=1$ to $n$}
\State $\mathbf{R}' \leftarrow \mathbf{R}$
\State $A \leftarrow \sum_{\mathbf{R}''} S^{fn}(\mathbf{R}'',\mathbf{R}')$, where $S^{fn}(\mathbf{R},\mathbf{R}')$ is defined in Eq.~(\ref{eq:Sfn}). 
\State Generate $\mathbf{R}$ according to the probability distribution $A^{-1}S^{fn}(\mathbf{R},\mathbf{R}')$. 
\State $w \leftarrow wA$
\EndFor
\State $(\phi_{l},w_{l}) \leftarrow (\mathbf{R},w)$
\State Compute the local energy $E^{\mathrm{loc}}(\phi_{l})$ according to Eq.~(\ref{eq:Eloc}). 
\EndFor
\State Output the energy $E\leftarrow \frac{\sum_{l=1}^N w_{l} E^{\mathrm{loc}}(\phi_{l})}{\sum_{l=1}^N w_{l}}$. 
\end{algorithmic}
}
\end{algorithm}
\end{minipage}
\end{figure*}

Given the basis, the imaginary-time Green's function reads 
\begin{align}
G(\mathbf{R},\mathbf{R}',\Delta\beta) = \bra{\mathbf{R}}e^{-\Delta\beta H}\ket{\mathbf{R}'}.
\end{align}
Considering the first-order expansion, we define 
\begin{align}
F(\mathbf{R},\mathbf{R}') \equiv \bra{\mathbf{R}}(\openone - \Delta\beta H)\ket{\mathbf{R}'}.
\end{align}
In the limit $n\rightarrow\infty$, $e^{-n\Delta\beta H}\ket{\Psi_I}$ converges to the ground state if $\ket{\Psi_I}$ has a finite overlap with the ground state. It is similar for the operator $F = \openone - \Delta\beta H$. Eigenstates of the Hamiltonian are eigenvectors of $F$, and the eigenvalue of the ground state is $1-\Delta\beta E_g$. If $1-\Delta\beta E_g$ is the largest absolute eigenvalue, $F^n\ket{\Psi_I}$ also converges to the ground state in the limit $n\rightarrow\infty$. If $\Delta\beta\norm{H}_2\leq 1$, $F$ is positive semi-definite, and the condition always holds. Note that sometimes it is helpful to add a proper constant to the Hamiltonian, i.e.~replace $H$ with $H-E_0$, where $E_0$ is the constant. 

According to $F^n\ket{\Psi_I}$, we compute the ground-state energy by evaluating  
\begin{align}
\frac{\bra{\Psi_T}HF^n\ket{\Psi_I}}{\bra{\Psi_T}F^n\ket{\Psi_I}} &=& \frac{\sum_{\mathbf{R}_0,\ldots,\mathbf{R}_n\in\mathcal{R}} \bra{\Psi_T}H\ket{\mathbf{R}_n}F(\mathbf{R}_n,\mathbf{R}_{n-1})\cdots F(\mathbf{R}_1,\mathbf{R}_0)\braket{\mathbf{R}_0}{\Psi_I}}{\sum_{\mathbf{R}_0,\ldots,\mathbf{R}_n\in\mathcal{R}} \braket{\Psi_T}{\mathbf{R}_n}F(\mathbf{R}_n,\mathbf{R}_{n-1})\cdots F(\mathbf{R}_1,\mathbf{R}_0)\braket{\mathbf{R}_0}{\Psi_I}} \notag \\
&=& \frac{\sum_{\mathbf{R}_0,\ldots,\mathbf{R}_n\in\mathcal{R}} E^{\mathrm{loc}}(\mathbf{R}_n) S(\mathbf{R}_n,\mathbf{R}_{n-1})\cdots S(\mathbf{R}_1,\mathbf{R}_0) \braket{\Psi_T}{\mathbf{R}_0}\braket{\mathbf{R}_0}{\Psi_I}}{\sum_{\mathbf{R}_0,\ldots,\mathbf{R}_n\in\mathcal{R}} S(\mathbf{R}_n,\mathbf{R}_{n-1})\cdots S(\mathbf{R}_1,\mathbf{R}_0) \braket{\Psi_T}{\mathbf{R}_0}\braket{\mathbf{R}_0}{\Psi_I}}.
\end{align}
In the last line, a similarity transformation is applied to $F(\mathbf{R},\mathbf{R}')$, and  
\begin{align}
S(\mathbf{R},\mathbf{R}') \equiv F(\mathbf{R},\mathbf{R}')\frac{\braket{\Psi_T}{\mathbf{R}}}{\braket{\Psi_T}{\mathbf{R}'}}.
\end{align}

In the approach introduced by van Bemmel {\it et al.}, The fixed-node Hamiltonian reads 
\begin{align}
\bra{\mathbf{R}}H^{fn}\ket{\mathbf{R}'} &\equiv & \left\{\begin{array}{ll}
\bra{\mathbf{R}}H\ket{\mathbf{R}'}+(1+\gamma)V^{sf}_{\mathbf{R}}, & \text{for }\mathbf{R}=\mathbf{R}', \\
\bra{\mathbf{R}}H\ket{\mathbf{R}'}, & \text{for }\mathbf{R}\neq\mathbf{R}'\text{ and }S(\mathbf{R},\mathbf{R}')\geq 0, \\
-\gamma\bra{\mathbf{R}}H\ket{\mathbf{R}'}, & \text{for }\mathbf{R}\neq\mathbf{R}'\text{ and }S(\mathbf{R},\mathbf{R}')<0, \\
\end{array}\right.
\end{align}
where 
\begin{align}
V^{sf}_{\mathbf{R}} \equiv \sum_{\mathbf{R}''\neq\mathbf{R}~:~G_T(\mathbf{R},\mathbf{R}'')<0} \bra{\mathbf{R}}H\ket{\mathbf{R}''} \frac{\braket{\Psi_T}{\mathbf{R}''}}{\braket{\Psi_T}{\mathbf{R}}}.
\end{align}
is the sign-flip potential at $\mathbf{R}$. Accordingly, 
\begin{align}
S^{fn}(\mathbf{R},\mathbf{R}') \equiv \bra{\mathbf{R}}(\openone - \Delta\beta H^{fn})\ket{\mathbf{R}'} \frac{\braket{\Psi_T}{\mathbf{R}}}{\braket{\Psi_T}{\mathbf{R}'}},
\label{eq:Sfn}
\end{align}
which is non-negative when $\gamma\geq 0$. In numerical simulations in this paper, we take $\gamma = 0$. Essential steps of fn-GFMC are given in Algorithm~\ref{alg:fn-GFMC}.

\section{Auxiliary-filed Monte Carlo}
\label{app:AFMC}

\subsection{Theory}

\begin{figure*}
\begin{minipage}{\linewidth}
\begin{algorithm}[H]
{\small
\begin{algorithmic}[1]
\caption{{\small Phaseless auxiliary-filed Monte Carlo algorithm.}}
\label{alg:ph-AFMC}
\Statex
\State Input $H$ [in the quadratic form of Eq.~(\ref{eq:quadratic})], $E_0$, $\Psi_I$, $\Psi_T$, $N$, $n$, $\Delta \beta$. 
\For{$l=1$ to $N$}
\State $\ket{\phi_{l,0}}\leftarrow \ket{\Psi_I}$
\State $w_{l,0}\leftarrow 1$
\Comment Initialise the walker. 
\For{$k=1$ to $n$}
\For{$j=1$ to $L$}
\State $\bar{x}_j \leftarrow -\sqrt{\Delta\beta}\frac{\bra{\psi_T}A_j\ket{\phi_{l,k-1}}}{\braket{\psi_T}{\phi_{l,k-1}}}$
\EndFor
\State Generate $\mathbf{x}$ according to the probability distribution $p(\mathbf{x})$ in Eq.~(\ref{eq:px}). 
\State $\ket{\phi_{l,k}} \leftarrow B(\mathbf{x}-\bar{\mathbf{x}})\ket{\phi_{l,k-1}}$, where $B(\mathbf{x}-\bar{\mathbf{x}})$ is given in Eq.~(\ref{eq:Bx}). 
\Comment Update the state. 
\State $E_{l,k-1} \leftarrow \frac{\bra{\psi_T}H\ket{\phi_{l,k-1}}}{\braket{\psi_T}{\phi_{l,k-1}}}$
\Comment Compute the local energy. 
\State $\theta \leftarrow \arg\left(\frac{\braket{\psi_T}{\phi_{l,k}}}{\braket{\psi_T}{\phi_{l,k-1}}}\right)$
\Comment Compute the phase. 
\State $I\leftarrow e^{-\Delta\beta \left(\Re E_{l,k-1}-E_0\right)}\times \max\left(0,\cos\theta\right)$
\State $w_{l,k}\leftarrow I\times w_{l,k-1}$
\Comment Update the weight. 
\EndFor
\State $E_{l,n} \leftarrow \frac{\bra{\psi_T}H\ket{\phi_{l,n}}}{\braket{\psi_T}{\phi_{l,n}}}$
\Comment Compute the final local energy.
\EndFor
\State Output the energy $E\leftarrow \frac{\sum_{l=1}^N w_{l,n} E_{l,n}}{\sum_{l=1}^N w_{l,n}}$. 
\end{algorithmic}
}
\end{algorithm}
\end{minipage}
\end{figure*}

For many fermion models, such as molecules, the Hamiltonian can be expressed in the form 
\begin{align}
H = A_0 - \frac{1}{2} \sum_{j=1}^{L} A_j^2,
\label{eq:quadratic}
\end{align}
where $A_j = \mathbf{A}_{j,\openone}\openone + \sum_{k,q}\mathbf{A}_{j,p,q}a_p^\dag a_q$ are one particle operators, and $a_q$ is the fermion annihilation operator of the $q$th spin orbital. The evolution time is divided into $n$ small time steps, i.e.~$\beta = n\Delta\beta$. Using the Trotter-Suzuki formula and Hubbard-Stratonovich transformation, the time evolution operator of each time step is rewritten as 
\begin{align}
e^{-\Delta\beta (H-E_0)} = \int d\mathbf{x}p(\mathbf{x}-\bar{\mathbf{x}})B(\mathbf{x}-\bar{\mathbf{x}}) + O(\Delta\beta^2),
\end{align}
where $\mathbf{x} = (x_1,x_2,\ldots,x_L)$ denotes the auxiliary-filed, 
\begin{align}
p(\mathbf{x}) \equiv (2\pi)^{-\frac{L}{2}} e^{-\frac{\abs{\mathbf{x}}^2}{2}}
\label{eq:px}
\end{align} 
is the normal distribution, 
\begin{align}
B(\mathbf{x}) \equiv \exp\left(-\Delta\beta (A_0-E_0)+\sqrt{\Delta\beta}\sum_{j=1}^{L}x_jA_j\right)
\label{eq:Bx}
\end{align}
are one-particle propagators, and $E_0$ is a constant taken a value close to the ground-state energy. Then we compute the ground-state energy by evaluating 
\begin{align}
&\frac{\bra{\Psi_T}He^{-n\Delta\beta (H-E_0)}\ket{\Psi_I}}{\bra{\Psi_T}e^{-n\Delta\beta (H-E_0)}\ket{\Psi_I}} \notag \\
&= \frac{\int d\mathbf{x}_1\cdots d\mathbf{x}_n p(\mathbf{x}_1-\bar{\mathbf{x}}_1)\cdots p(\mathbf{x}_n-\bar{\mathbf{x}}_n) \bra{\Psi_T}HB(\mathbf{x}_n-\bar{\mathbf{x}}_n)\cdots B(\mathbf{x}_1-\bar{\mathbf{x}}_1)\ket{\Psi_I} + O(n\Delta\beta^2)}{\int d\mathbf{x}_1\cdots d\mathbf{x}_n p(\mathbf{x}_1-\bar{\mathbf{x}}_1)\cdots p(\mathbf{x}_n-\bar{\mathbf{x}}_n) \bra{\Psi_T}B(\mathbf{x}_n-\bar{\mathbf{x}}_n)\cdots B(\mathbf{x}_1-\bar{\mathbf{x}}_1)\ket{\Psi_I} + O(n\Delta\beta^2)} \notag \\
&= \frac{\int d\mathbf{x}_1\cdots d\mathbf{x}_n p(\mathbf{x}_1)\cdots p(\mathbf{x}_n) E^{\mathrm{loc}}(\Psi_n) I(\mathbf{x}_n,\bar{\mathbf{x}}_n,\Psi_{n-1})\cdots I(\mathbf{x}_1,\bar{\mathbf{x}}_1,\Psi_{0})}{\int d\mathbf{x}_1\cdots d\mathbf{x}_n p(\mathbf{x}_1)\cdots p(\mathbf{x}_n) I(\mathbf{x}_n,\bar{\mathbf{x}}_n,\Psi_{n-1})\cdots I(\mathbf{x}_1,\bar{\mathbf{x}}_1,\Psi_{0})} + O(n\Delta\beta^2).
\end{align}
where the importance function is 
\begin{align}
I(\mathbf{x},\bar{\mathbf{x}},\phi) \equiv \frac{\bra{\psi_T}B(\mathbf{x}-\bar{\mathbf{x}})\ket{\phi}}{\braket{\psi_T}{\phi}}e^{\mathbf{x}\cdot\bar{\mathbf{x}}-\frac{\bar{\mathbf{x}}\cdot\bar{\mathbf{x}}}{2}},
\end{align}
\begin{align}
\ket{\Psi_k} \equiv B(\mathbf{x}_k-\bar{\mathbf{x}}_k)\cdots B(\mathbf{x}_1-\bar{\mathbf{x}}_1)\ket{\Psi_I},
\end{align}
and $\ket{\Psi_0} \equiv \ket{\Psi_I}$. We take each $\bar{\mathbf{x}}$ according to 
\begin{align}
\bar{x}_j = -\sqrt{\Delta\beta}\frac{\bra{\psi_T}A_j\ket{\phi}}{\braket{\psi_T}{\phi}},
\label{eq:xbar}
\end{align}
and $\ket{\phi} = \ket{\Psi_{k-1}}$ for $\bar{\mathbf{x}}_k$. In ph-AFMC, we approximate the importance function with a non-negative number. Specifically, in the phaseless approximation, we take 
\begin{align}
 I(\mathbf{x},\bar{\mathbf{x}},\phi)\approx e^{-\Delta\beta \left[\Re E^{\mathrm{loc}}(\phi)-E_0\right]}  \times \max\left(0,\cos\left(\arg\left(\frac{\bra{\psi_T}B(\mathbf{x}-\bar{\mathbf{x}})\ket{\phi}}{\braket{\psi_T}{\phi}}\right)\right)\right).
\label{eq:Iapp}
\end{align}
Essential steps of ph-AFMC are given in Algorithm~\ref{alg:ph-AFMC}. 

\subsection{Numerical simulation results}
\label{app:afmc:simulation}

\begin{figure}[tbp]
\begin{center}
\includegraphics[width=0.6\linewidth]{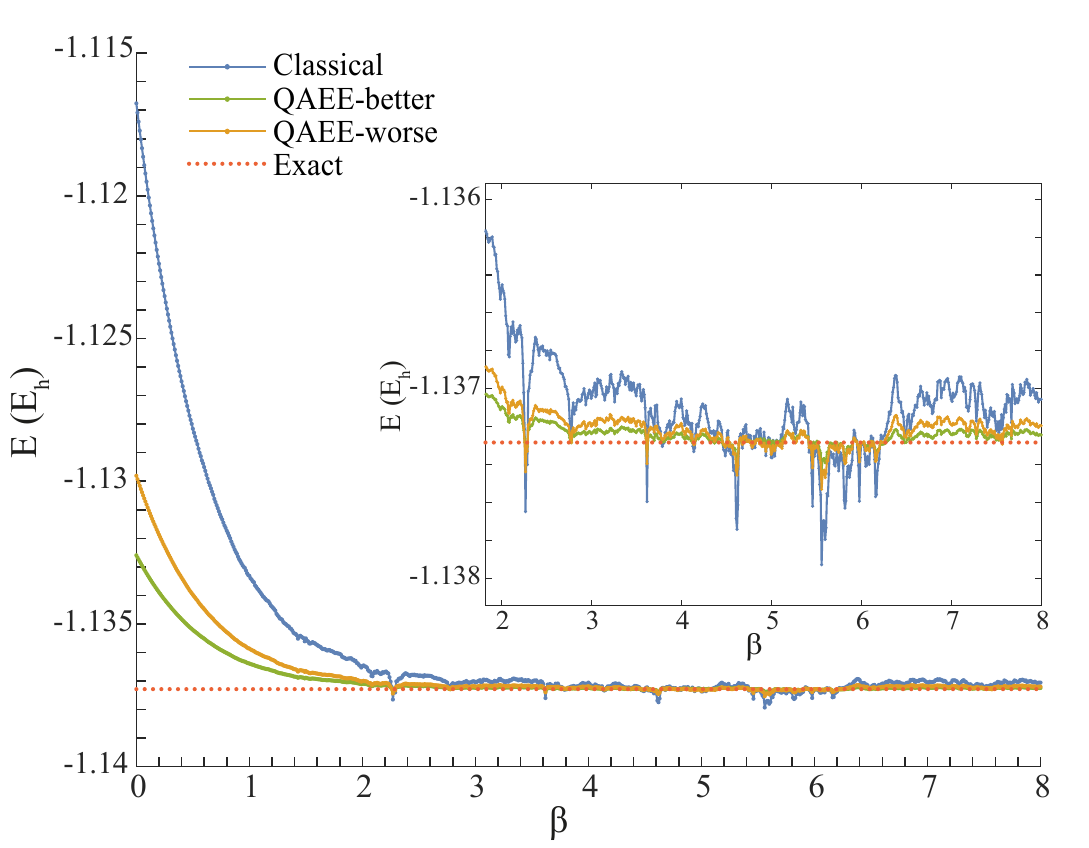}
\caption{
The energy ($\mathrm{mE_h}$) of a $\mathrm{H_2}$ molecule computed using the classical and the quantum-assisted energy evaluation ph-AFMC algorithms. We take $N=2\times 10^5$ and $\Delta\beta=0.01$. The yellow and green curves represent the cases with lower and higher state fidelities (with respect to the true ground state) of quantum computing, respectively. The exact ground-state energy is plotted as the red dotted curve. A zoomed-in image of the range of $\beta$ from around 2 to 4 is shown as the inserted graph for better illustration. 
}
\label{fig:plot_afmc}
\end{center}
\end{figure}

\begin{table}[H]
\centering
\resizebox{0.45\columnwidth}{!}{\begin{tabular}{|c|c|c|}
\hline
 & $H_2$ & $H_4$ \\ \hline
\begin{tabular}{c}Classical algorithm\\\end{tabular} & 0.211 & 0.224 \\ \hline
\begin{tabular}{c}QAEE \end{tabular} & 0.0794 & 0.0796 \\ \hline
\end{tabular}}
\caption{
Errors in the ground-state energy ($\mathrm{mE_h}$) of a $\mathrm{H_2}$ and a $\mathrm{H_4}$ linear chains computed with the classical and quantum-assisted ph-AFMC algorithms. For all cases we take $N=2\times 10^5$. 
}
\label{table:QAEEAFMC}
\end{table}

In this section, we present supplementary numerical results of the classical and quantum-assisted AFMC algorithms. A $\mathrm{H_4}$ linear chain and a $\mathrm{H_2}$ Hydrogen molecule are considered, and for both cases, the interatomic distance is $0.74 \mathring{A}$. Taking the STO-3G basis, we take the Hartree-Fock state as CCTS (as it is commonly used), which has the fidelity of $98.73\%$ and $97.41\%$ with respect to the ground state of $\mathrm{H_2}$ and $\mathrm{H_4}$, corresponding to an error of $20.525\mathrm{mE_h}$ and $40.99\mathrm{mE_h}$ in the ground state energy, respectively. For the quantum-assisted algorithm, we use two UCCSD trial states for $\mathrm{H_2}$, with a state fidelity of $98.72\%$ and $99.41\%$, respectively, which correspond to an error of $12.001\mathrm{mE_h}$ and $4.733\mathrm{mE_h}$; for $\mathrm{H_4}$, we use a UCCSD trial state with a fidelity of $98.38\%$ and an error of $19.885\mathrm{mE_h}$. The phaseless approximation is used in the simulation to eliminate the sign problem. 

The result of the energy of a $\mathrm{H_2}$ molecule is shown in Fig.~\ref{fig:plot_afmc}. As $\beta$ increases, all the three curves stabilise to approach the exact ground state, while the blue curve (representing the result from the classical algorithm) is observed to have a larger bias and variance than the two other curves (representing results from the quantum-assisted algorithm with different trial states), indicating that the quantum-assisted algorithm generates a better result. Similar results are observed for $\mathrm{H_4}$, as summarised in Table~\ref{table:QAEEAFMC}, which shows the error in the ground state energy. Compared with the results for VMC and fn-GFMC, we see the error from both the classical and quantum-assisted ph-AFMC algorithms is smaller, as is usually the case for classical algorithms; that’s because in AFMC, the walker state is a general Slater determinant close to the ground state and the local energy fluctuates around the actual energy. On the other hand, such a feature also makes AFMC more challenging to implement for both classical and quantum-assisted algorithms. 

\section{Detailed derivation in amplitude estimation}

\subsection{The vacuum reference method}
\label{app:vacuum_reference}

In this section we show the circuit in Fig.~\ref{fig:circuits}(b) can compute the real and imaginary parts of the amplitude $\langle\Psi_T|\phi\rangle$. 

In the circuit, the first Hadamard gate prepares the state $\ket{0}^{\otimes n_q}+\ket{1}^{\otimes n_q-1}$ and the following CNOT gates transform the state into 
\begin{align}
    \ket{S}=\frac{1}{\sqrt{2}}(\ket{\Phi_r}+\ket{HF}).
\end{align}
The first qubit is measured in the $X$ or $Y$ basis, while all the other qubits are measured in $\ket{0}$ state. Therefore, the circuit effectively measures the operator $\widetilde{X}=X\otimes\ket{0}\bra{0}^{\otimes n_q-1}$:
\begin{align}
\langle \widetilde{X}\rangle=\langle S|V_{\phi}^{\dag}V_TA \widetilde{X}A^\dag V_T^{\dag}V_\phi|S\rangle,
\label{app:eq:M}
\end{align}
where $A^\dag$ refers to those CNOT gates before measurement. $A$ satisfies $A\ket{0}^{\otimes n_q}=\ket{0}^{\otimes n_q}$ and $A\ket{1}\otimes\ket{0}^{\otimes n_q-1}=\ket{HF}$, thus
\begin{align}
A\widetilde{X}A^\dag = \ket{HF}\bra{\Phi_r}+\ket{\Phi_r}\bra{HF}.
\label{app:eq:ama}
\end{align}

As we know $\langle \Phi_r|V_T^{\dag}V_\phi|\Phi_r\rangle = 1$ and $\langle HF|V_T^{\dag}V_\phi|\Phi_r\rangle = \langle \Phi_r|V_T^{\dag}V_\phi|HF\rangle =0$, taking Eq.~\ref{app:eq:ama} into Eq.~\ref{app:eq:M}, Eq.~\ref{app:eq:M} thus becomes
\begin{align}
\langle \widetilde{X}\rangle= & \frac{1}{2}\left(\langle HF|V_{\phi}^{\dag}V_T|HF\rangle + \langle HF|V_T^{\dag}V_{\phi}|HF\rangle\right) \\
=&\Re\left(\langle HF|V_T^{\dag}V_\phi|HF\rangle\right)=\Re\left(\langle\Psi_T|\phi\rangle\right).
\end{align}
Similarly, we can find $\langle\widetilde{Y}\rangle = \Im\left(\langle\Psi_T|\phi\rangle\right)$. Thus we obtain the amplitude $\langle\Psi_T|\phi\rangle$.

\subsection{The Hartree-Fork reference method}
\label{app:hartree_fork}
We can write the trial state $\ket{\Psi_T}$ as
\begin{align}
\ket{\Psi_T} = \alpha_\mathbf{R}\ket{\mathbf{R}}+\alpha_{HF}\ket{HF}+\alpha_0\ket{\Psi_0},
\label{app:eq:psi_t}
\end{align}
where $\ket{\Psi_0}$ is a state orthogonal to $\ket{\mathbf{R}}$ and $\ket{HF}$. So the amplitude to find is $\langle\Psi_T|\mathbf{R}\rangle=\alpha_\mathbf{R}$.

The circuit in Fig.~\ref{fig:circuits}(c) measures $\widetilde{X}$ or $\widetilde{Y}$. We take $\widetilde{X}$ as the example. The circuit evaluates 
\begin{align}
\langle \widetilde{X}\rangle=\langle 0|U_T^\dag U_R \widetilde{X}U_R^\dag U_T|0\rangle=\langle \Psi_T|U_R \widetilde{X}U_R^\dag|\Psi_T\rangle.
\label{app:eq:ut}
\end{align}

Considering $\ket{HF} = U_{\mathbf{R}}\ket{0}^{\otimes n_q}$, $\ket{\mathbf{R}} = U_{\mathbf{R}}\ket{1}\otimes\ket{0}^{\otimes(n_q-1)}$, we take Eq.~\ref{app:eq:psi_t} into Eq.~\ref{app:eq:ut} and obtain
\begin{align}
\langle \widetilde{X}\rangle =\alpha_{\mathbf{R}}\alpha_{HF}^*+\alpha_{\mathbf{R}}^*\alpha_{HF}.
\end{align}

As we assume $\braket{\Psi_T}{HF}$ is positive, $\alpha_{HF}$ is real, thus 
\begin{align}
\Re(\alpha_{\mathbf{R}}) = \frac{\langle \widetilde{X}\rangle}{2\alpha_{HF}}.
\end{align}

Similarly we can find 
\begin{align}
\Im(\alpha_{\mathbf{R}}) = \frac{\langle \widetilde{Y}\rangle}{2\alpha_{HF}},
\end{align}
thus
\begin{align}
\alpha_{\mathbf{R}} = \frac{\langle \widetilde{X}\rangle+i\langle \widetilde{Y}\rangle}{2\alpha_{HF}}.
\end{align}

\end{document}